\documentclass[10pt,dvipsnames,aps,prl,nobibnotes,twocolumn,nolongbibliography,floatfix]{revtex4-2}
    \setcitestyle{super}
    \makeatletter
        \def\input@path{{src/}{tikz/}}
    \makeatother
    \usepackage{graphicx}
    \usepackage{float}
    \usepackage{enumitem}
        \newlist{ipclist}{enumerate}{1}
        \setlist[ipclist,1]{label=\bf\Alph*}
    \usepackage[x11names, table, svgnames]{xcolor}
        \definecolor{hypercitecolor}{HTML}{4287f5}
    \usepackage[T1]{fontenc}
    \usepackage[utf8]{inputenc}

    \usepackage{mathtools}
    \usepackage{amsthm}

    \usepackage{amssymb}
    \usepackage{pdfpages}
    \usepackage{array}
    \usepackage{mathrsfs}
    \usepackage{bm}
    \usepackage{xparse}
    \usepackage{siunitx}
    \usepackage{stackrel}
    \usepackage{nicefrac}
    \newcommand{\new}[1]{{\color{black} {#1}}}

    \usepackage[acronym]{glossaries}
        \newacronym{lrg}{LRG}{Laplacian Renormalization Group}
        \newacronym{sl}{SL}{Signed Laplacian}
        \newacronym{si}{SI}{Supplementary Information}
        \newacronym{sg}{SG}{Spin Glass}
        \newacronym{rbim}{RBIM}{Random Bond Ising Model}
        \newacronym{rsb}{RSB}{Replica Symmetry Breaking}
        \newacronym{ea}{EA}{Edwards-Anderson}
        \newacronym{gc}{GC}{Giant Component}
        \newacronym{sw}{SW}{Small World}
        \newacronym{er}{ER}{Erd\H{o}s-R\'enyi}
        \newacronym{rg}{RG}{Renormalization Group}
\let\de\partial
\let\quenchavg\overline

\let\vec\bm
\let\adjacencymatrixsymb A
\let\bigOnotationsymb O
\let\continuousfieldsymb\psi
\let\couplingmatrixsymb J
\let\critsymb c
\let\degreematsymb D
\let\densitymatrixsymb \rho
\let\diracdeltasymb \delta
\let\discretefieldsymbol \phi
\let\edgesetsymb E
\let\entropysymb S
\let\functionaldifferentialsymbol D
\let\graphsymb G

\let\hamiltoniansymb H
\let\laplaciansymb L
\let\magnetizationsymbol m
\let\nodedegreesymb k
\let\partitionfunctionsymb Z
\let\probabilitysymb P

\let\sinkoperatorsymb S
\let\spinsymb\sigma
\let\vertexsymb V
\let\weightedadjacencymatrixsymb W

\newcommand{\dilutedshrt}{dil}

\newcommand{\rewiringshrt}{rew}
\newcommand{\signshrt}{sign}

\newcommand{\adj}{\adjacencymatrixsymb}
\newcommand{\cfield}{\continuousfieldsymb}
\newcommand{\crit}{{\rm \critsymb}}
\newcommand{\ddelta}{\diracdeltasymb}
\newcommand{\dfield}{\discretefieldsymbol}
\newcommand{\degm}{\degreematsymb}
\newcommand{\dil}{{\rm \dilutedshrt}}
\newcommand{\dm}{\hat{\densitymatrixsymb}}

\newcommand{\fundiff}{\mathcal{D}}

\newcommand{\ham}{\mathscr{\hamiltoniansymb}}
\newcommand{\jij}{\couplingmatrixsymb}
\newcommand{\lapl}{\laplaciansymb}

\newcommand{\ndegree}{\nodedegreesymb}

\newcommand{\partf}{\partitionfunctionsymb}
\newcommand{\prob}{\probabilitysymb}
\newcommand{\rew}{{\rm \rewiringshrt}}
\newcommand{\sdegm}{\bar{\degm}}

\newcommand{\Slapl}{\bar{\lapl}}
\newcommand{\spin}{\spinsymb}

\DeclareMathOperator{\sign}{\signshrt}

\DeclareMathOperator{\Lapl}{\nabla^2}

\DeclareMathOperator{\Tr}{Tr}
\DeclarePairedDelimiter{\abs}{|}{|}
\DeclarePairedDelimiter{\avg}{\langle}{\rangle}

\DeclarePairedDelimiter{\ket}{|}{\rangle}

\DeclarePairedDelimiterX{\braket}[2]{\langle}{\rangle}{#1\,\delimsize\vert\,\mathopen{}#2}

\NewDocumentCommand{\dd}{s o m}{
  \IfBooleanTF{#1} 
    {\mathrm{d}#3} 
    {\IfNoValueTF{#2} 
      {\mathrm{d}#3} 
      {\mathrm{d}^{#2}#3} 
    }
}

\NewDocumentCommand{\dv}{s o m m}{
  \IfBooleanTF{#1} 
    {{\mathrm{d}#3}/{\mathrm{d}#4}} 
    {\IfNoValueTF{#2} 
      {\frac{\mathrm{d}#3}{\mathrm{d}#4}} 
      {\frac{\mathrm{d}^{#2}#3}{\mathrm{d}#4^{#2}}} 
    }
}

\NewDocumentCommand{\pdv}{s o m m}{
  \IfBooleanTF{#1} 
    {{\partial #3}/{\partial #4}} 
    {\IfNoValueTF{#2} 
      {\frac{\partial #3}{\partial #4}} 
      {\frac{\partial^{#2} #3}{\partial #4^{#2}}} 
    }
}

    \usepackage[caption=false]{subfig}
    \usepackage{tikz}
        \usetikzlibrary{positioning}
        \usetikzlibrary{fit}
        \tikzset{graph node/.style={draw, circle}}
        \tikzset{dgraph node/.style={draw, densely dashed, circle}}
    
    \PassOptionsToPackage{hyphens}{url}
    \usepackage{hyperref}
        \hypersetup{
            colorlinks,
            citecolor=blue,
            urlcolor=blue,
            linkcolor=blue,
        }
\makeatletter
    \renewcommand\paragraph{%
      \@startsection{paragraph}{4}{\z@}%
        {3.25ex \@plus1ex \@minus .2ex}%
        {-1em}%
        {\normalfont\normalsize\bfseries}%
    }
    \newcommand{\phantomlabel}[2]{
        \protected@write\@auxout{}{
            \string\newlabel{#2}{
                {\@currentlabel#1}{\thepage}
                {\@currentlabel#1}{#2}{}
            }
        }
        \hypertarget{#2}{}
    }
    \NewDocumentCommand{\FigRef}{m}{%
      \ifvmode Figure~\ref{#1}%
      \else Fig.~\ref{#1}%
      \fi
    }
    \renewcommand{\eqref}[1]{Eq.~(\ref{#1})}

\AtBeginDocument{\let\LS@rot\@undefined}
\makeatother
\begin{document}
    \title{Topological Symmetry Breaking in Antagonistic Dynamics}
    \author{Giulio Iannelli$^{1, 2}$}
    \author{Pablo Villegas$^{1, 3}$}
        \email{pablo.villegas@cref.it}
    \author{Tommaso Gili$^{4, 1}$}
    \author{Andrea Gabrielli$^{1, 5}$}
    \affiliation{$^1$Enrico Fermi Research Center (CREF), Via Panisperna 89A, 00184, Rome, Italy}
    \affiliation{$^2$Dipartimento di Fisica, Universit\`a di Roma ``Tor Vergata'', 00133 Rome, Italy}
    \affiliation{$^3$Instituto Carlos I de F\'isica Te\'orica y Computacional, Universidad de Granada, Granada, Spain}
    \affiliation{$^4$Networks Unit, IMT Scuola Alti Studi Lucca, Piazza San Francesco 15, 55100- Lucca, Italy.}
    \affiliation{$^5$Dipartimento di Ingegneria Civile, Informatica e delle Tecnologie Aeronautiche, Universit\`a degli Studi ``Roma Tre'', Via Vito Volterra 62, 00146, Rome, Italy}
    \maketitle
\textbf{A dynamic \emph{concordia discors}, a finely tuned equilibrium between opposing forces, is hypothesized to drive historical transformations \cite{gramsci_quaderni}. Similarly, a precise interplay of excitation and inhibition, the 80:20 ratio, is at the basis of the normal functionality of neural systems \cite{Xue2014,Harris2013,Froemke2007}. In artificial neural networks, reinforcement learning allows for fine-tuning internal signed connections, optimizing adaptive responses to complex stimuli, and ensuring robust performance \cite{Hopfield1982,AGS1985,Levine2024}. 
Engineered systems with antagonistic interactions remain comparatively unexplored, particularly because their emergent phases are closely linked with frustration mechanisms in the hosting network \cite{mezard2023,parisi2023nobel}. In this context, the spin glass theory has shown how an apparently uncontrollable non-ergodic chaotic behavior arises from the complex interplay of competing interactions and frustration among units \cite{bray1987,castellani2005spin}, leading to multiple metastable states preventing the system from exploring all accessible configurations over time \cite{mezard1984}. \new{Here, we show how topology constrains dynamics in systems with antagonistic interactions.} We make use of the signed Laplacian operator to demonstrate how fundamental \emph{topological defects} in lattices and networks percolate, shaping the geometrical arena and complex energy landscape of the system. This unveils novel, highly robust multistable phases and establishes deep connections with spin glasses when thermal noise is considered, providing a natural topological and algebraic \new{description of emergent multistability and non-ergodicity in frustrated systems.}}



Competing interactions introduce intrinsic incompatibilities between fundamental dynamical interactions and the elementary scales of the underlying lattice geometry \cite{chowdhury2014spin}, shaping the collective behavior of the system. This gives rise to new fundamental phenomena and leads to intriguing effects, such as the formation of exotic states like spin ice \cite{Moessner2006,Harris1999}, spin liquids \cite{Anderson1987}, and spin glasses that have tantalized physicists for the last 40 years \cite{charbonneau2023spin,mezard2023}. The latter, the paradigm of disordered magnetic systems, are characterized by competing interactions among their constituent spins, leading to an intricate and unpredictable \emph{dynamical} and statistical configurational behavior \cite{castellani2005spin}. Signed interactions induce frustration and play a crucial role in describing the complex dynamics of these systems. 

The thorough study of spin glasses opened the door to one of the most elegant proposals in the field of modern theoretical physics: the replica trick and the \emph{replica symmetry breaking} (RSB), which offers a comprehensive understanding of their non-ergodic thermodynamic properties \cite{mezard1987spin,parisi1979infinite,parisi1983} in the mean-field scenario. Solid efforts have been made to detect the nature of the spin glass phase \cite{mezard1984,bray1987,nishimoribook}. Therefore, significant progress in both analytical and numerical aspects has been made: the discovery of temperature and disorder chaos \cite{katzgraber2007}, stochastic stability \cite{aizenman1998}, metastates \cite{amit1985}, or the rigorous analysis of short-range spin glasses \cite{marinari2000,fisher1986}. In a pioneering work, Amit et al. established the rigorous connection between neural networks and Ising spin glasses, highlighting that both systems share an energy landscape with numerous metastable states and valleys that enable the delocalized storage of patterns \cite{amit1985}. 

However, several essential questions remain unanswered. Little is known about the number of pure ground states or the nature of broken symmetry in short-range spin glasses \cite{Stein2013}. Moreover, determining the ground states of Ising spin glasses in three or more dimensions is an NP-hard problem \cite{Barahona1982}, which is closely related to many other hard combinatorial optimization problems \cite{Lucas2014}. Similarly, disentangling new ensembles of correlated disorder \cite{mezard2023}, that is, predicting spin glass states in networks, represents an open challenge that remains crucial. In arbitrary architectures, in turn, it has never been fully understood the interplay between topology and signed interactions nor modeled such a delicate balance to generate multiple local states. Specifically, understanding the impact of competing units, i.e., elementary defects, in regular lattices and complex networks is key to properly controlling quantum error-correcting codes \cite{Kitaev2006} and developing innovative brain-inspired devices, or even entirely new classes of structured artificial intelligence models that allow for more efficient and richer scenarios of machine-learning algorithms.
\newpage
\section*{Canonical formulation of signed architectures}
The multiscale structure of any homogeneous or heterogeneous signed network or lattice with antagonistic interactions can be analyzed using its corresponding Laplacian density matrix \cite{dedomenico2016spectral, InfoCore, LRG},
    \begin{equation}
        \dm(t) = \frac{e^{-t\Slapl}}{\partf(t)}
            \label{eq:density_matrix}
    \end{equation}
where \(\partf(t)=\Tr[e^{-t\Slapl}]\), \(\Slapl=\sdegm-\adj\) represents the \gls{sl} of the network \cite{kunegis2010spectral}, with \(\adj\) being the signed adjacency matrix and \(\sdegm\) the diagonal matrix of elements \((\sdegm)_{ij} = \delta_{ij}\sum_k\abs{\adj_{ik}}\). $\Slapl$ can be derived from the usual reordering of the discrete Landau-Ginzburg Hamiltonian \cite{Binney} and extends the Laplacian operator to signed architectures (see Methods).

\FigRef{fig:a:lattc2sq_panel} illustrates the \gls{sl} in terms of spectral embedding arguments, as originally developed in \cite{kunegis2010spectral} using the concept of \emph{antipodal proximity}. Ordinary Laplacian embedding relies on the non-trivial solutions of the Laplace equation \(\lapl \vec{x} = 0\) for placing vertices of a graph in Euclidean space. In this way, each vertex gets positioned at the weighted mean of its neighbors' coordinates, with weights determined by the connecting edges. In contrast, when using \(\Slapl \vec{x} = 0\) for embedding a signed graph, negatively connected neighbors contribute their coordinates with an opposite sign to the weighted sum determining the node placement. This makes nodes get closer to the antipodes of their oppositely tied neighbors. Moreover, since the \gls{sl} has been formally proven to be a positive semidefinite operator (see Methods), we can rigorously implement the recently introduced Laplacian framework \cite{LRG,InfoCore} to \emph{scan} different spatial scales when an arbitrary architecture contains antagonistic or \emph{frustrated interactions}. In particular, the heat capacity of a signed graph
    \begin{equation}
        C = -\dv{S}{(\log \tau)},
            \label{eq:C_dvSlogtau}
    \end{equation}
defined as the derivative of the graph entropy \cite{dedomenico2016spectral, InfoCore, LRG} \(S(t)=\Tr[\dm(t)\ln \dm(t)]\) (see Methods and \gls{si}) as a function of the (logarithm of the) diffusion time \(\tau\) is, therefore, a detector of transition points that reveals the intrinsic characteristic diffusion scales of the system. Indeed, each pronounced peak of \(C\) reveals a strong deceleration of the information diffusion, separating regions with strong communicability from the rest of the network where the diffusion slows down \cite{InfoCore, LRG}.

\FigRef{fig:b:lattc2sq_panel} shows a sketch of a 2D lattice with \(M\) edges, where a random fraction of negative edges, denoted by \(p=M^-/M\), is introduced, being \(M^-\) the total number of negative edges.  \FigRef{fig:c:lattc2sq_panel} shows the average heat capacity \(\quenchavg{C}\) for different values of \(p\) in 2D square lattices. In particular, we observe that the peak at short diffusion times disappears as the number of negative links increases, disrupting the lattice translational invariance. \new{Note that the point where \(\quenchavg{C}\) exhibits a peak at short times corresponds to the shortest resolvable topological interaction scale (by analogy with an ultraviolet cutoff), which in regular lattices coincides with the lattice spacing \cite{LRG,JSTAT,TBM}.} The existence of this peak, together with the subsequent plateau, is closely related to the existence of translational invariance in the network\cite{Poggialini,TBM}, which is lost at some specific value \(p_\crit\) when negative links gradually replace the positive ones. The probability distribution of the first maximum of \(C\) for different realizations of 2D lattices with variable levels of frustration is reported in \FigRef{fig:d:lattc2sq_panel}, exhibiting bistable behavior and a Maxwell-like point for \(p_\crit\approx0.1\). This symmetry breaking in the very topology plays, as discussed below, a key role in determining, altering, and disrupting dynamic phases and phase transitions.

\begin{figure}[!h]
    \centering
    \includegraphics[width=\linewidth]{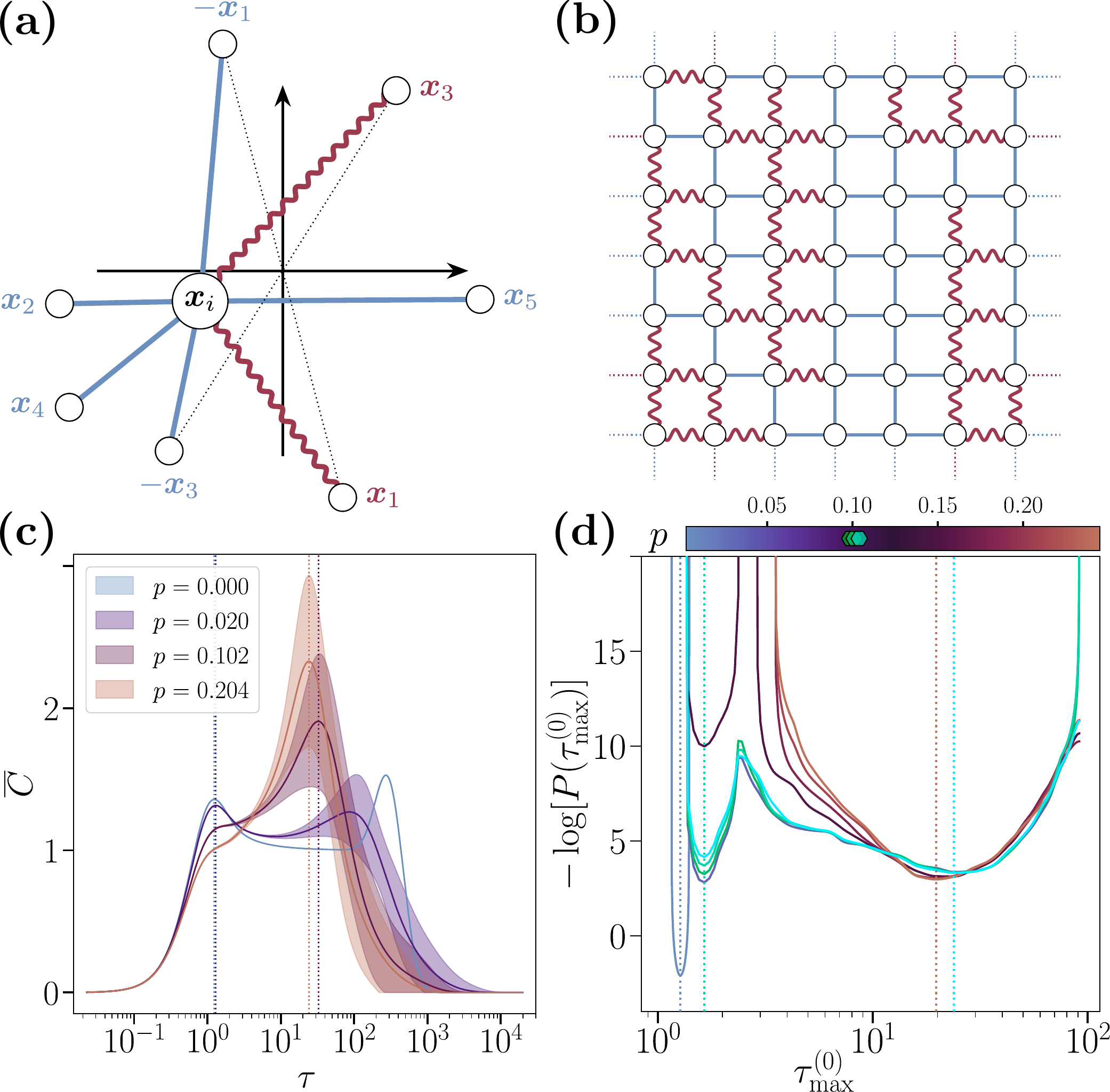}
    \caption{\textbf{Signed Laplacian. (a)} 
        The antipodal proximity concept in the 2D spectral embedding with the \gls{sl} \cite{kunegis2010spectral}. The position \(\vec{x}_i\) of the $i^{\mathrm{th}}$ node  is computed as the average position of the neighbors \(\vec{x}_j\) weighted by the adjacency index \(\adj_{ij}\). When \(\sign(\adj_{ij}) = -1\), the position \(\vec{x}_i\) is updated in the direction opposite to \(\vec{x}_j\). Black arrows correspond to coordinate axes. \textbf{(b)} Sketch of a 2D square lattice with random positive (blue links) and negative (red links) interactions. \textbf{(c)} Heat capacity, $C$, versus the temporal resolution parameter of the system, $\tau$, for different values of $p$ averaged over multiple disorder realizations (see legend). \textbf{(d)} Probability distribution, \(-\log[P(\tau^{(0)}_{\max})]\), for the position, $\tau^{(0)}_{\max}$, of the first detected maximum of $C$ at varying $p$. Its shape changes from a single minimum for low $p$ to a bistable configuration at $p_\crit$.}
    \label{fig:lattc2sq_panel}
    \phantomlabel{(a)}{fig:a:lattc2sq_panel}
    \phantomlabel{(b)}{fig:b:lattc2sq_panel}
    \phantomlabel{(c)}{fig:c:lattc2sq_panel}
    \phantomlabel{(d)}{fig:d:lattc2sq_panel}
\end{figure}
\begin{figure*}[!hbt]
    \centering
    \includegraphics[width=\textwidth]{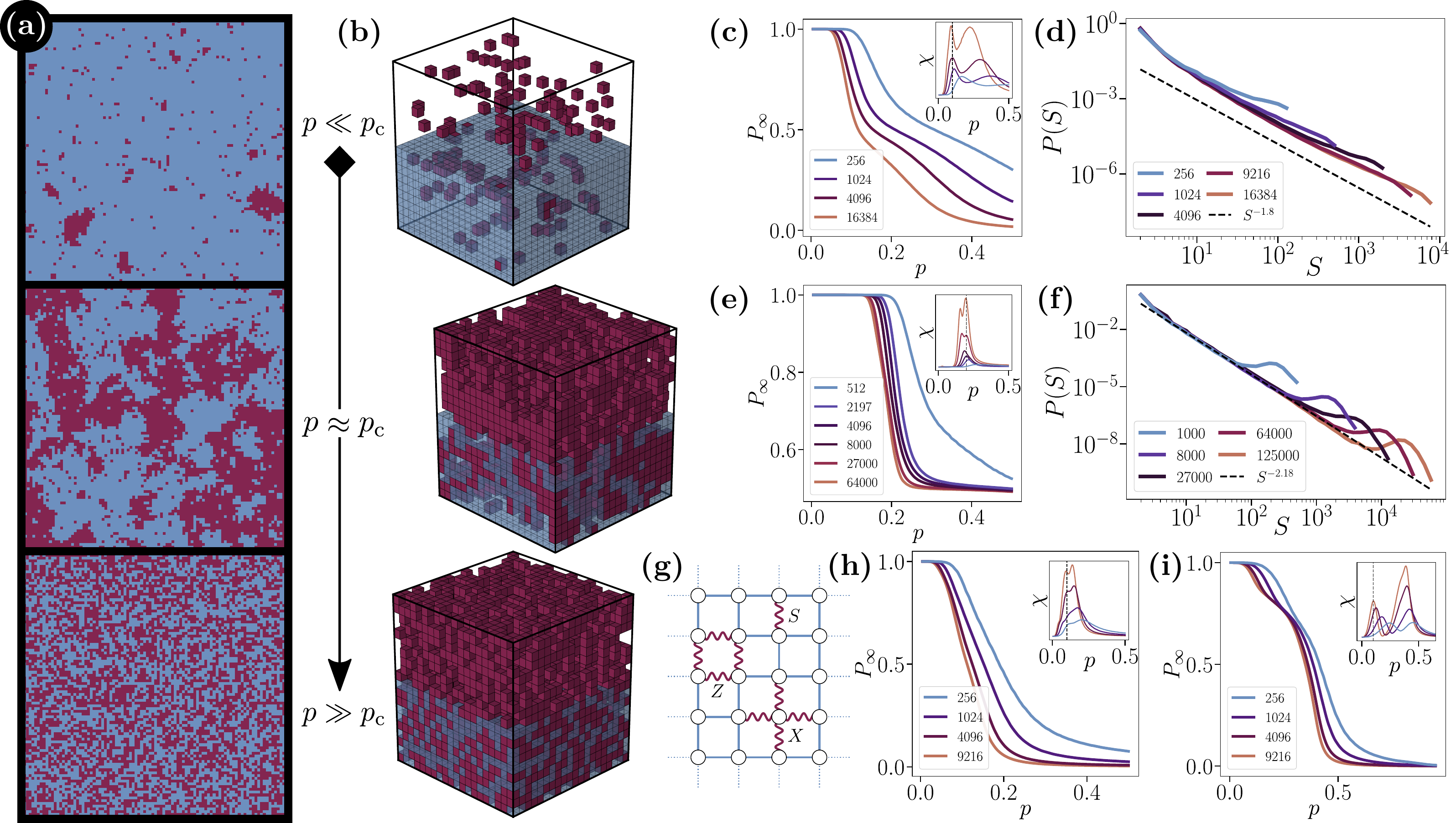}
    \caption{\textbf{Percolation of topological defects in regular structures. (a)} Partitioning of a square 2D lattice as indicated by the positive and negative signs of $\ket{\lambda_0}.$ A set of fractal clusters emerges beyond the critical value, $p_\crit$. $\textbf{(b)}$ The lower part of each figure shows the partitioning of a 3D lattice as indicated by the positive and negative signs of $\ket{\lambda_0}$ (blue and red nodes, respectively). The upper part of the figure enhances only the positive ones. \textbf{(c)}  Order parameter, $P_{\infty}$ versus the fraction of negative links, $p$ for a squared 2D lattice of variable size (see legend). Inset: Variance of the order parameter, $\chi$, as a function of the fraction of negative links, $p$. Note that $\chi$ diverges as the size of the system increases. Black dashed lines represent the expected value for the spin-glass transition at $T=0$. \textbf{(d)} Cluster size distribution for a 2D square lattice of different system sizes (see legend). The black dashed line represents the estimated Fisher exponent, $\tau\approx1.8$. \textbf{(e)} Order parameter, $P_{\infty}$ versus the fraction of negative links, $p$, for a 3D cubic lattice of variable size (see legend). Inset: Variance of the order parameter, $\chi$, as a function of the fraction of negative links, $p$. Note that $\chi$ diverges as the size of the system increases. Black dashed lines represent the expected value for the spin-glass transition at $T=0$. \textbf{(f)} Cluster size distribution for a cubic lattice of different system sizes (see legend). The black dashed line represents the estimated Fisher exponent, $\tau=2.18$. \textbf{(g)} Sketch of topological defects in a squared lattice: single links (S), Z defects representing the elementary cell, and X defects representing ``anti-nodes''. \textbf{(h)} $P_{\infty}$ versus the fraction of Z defects, $p$ for a squared 2D lattice of variable size (see legend).  Inset: Variance of the order parameter, $\chi$, as a function of the fraction of Z defects, $p$. The black dashed vertical line corresponds to $p_\crit=0.1$.  \textbf{(i)} $P_{\infty}$ versus the fraction of X defects, $p$ for a squared 2D lattice of variable size (see legend).  Inset: Variance of the order parameter, $\chi$, as a function of the fraction of X defects, $p$. The black dashed vertical line corresponds to $p_\crit=0.1$.}
    \label{fig:percolation_panel}
    \phantomlabel{(a)}{fig:a:percolation_panel}
    \phantomlabel{(b)}{fig:b:percolation_panel}
    \phantomlabel{(c)}{fig:c:percolation_panel}
    \phantomlabel{(d)}{fig:d:percolation_panel}
    \phantomlabel{(e)}{fig:e:percolation_panel}
    \phantomlabel{(f)}{fig:f:percolation_panel}
    \phantomlabel{(g)}{fig:g:percolation_panel}
    \phantomlabel{(h)}{fig:h:percolation_panel}
    \phantomlabel{(i)}{fig:i:percolation_panel}
\end{figure*}
%

\new{The heat capacity $C(\tau)$ serves as a spectral telescope: short times $\tau$ probe the high-energy (local) defects, while long times probe the low-energy (global) topological structure. Thus, anomalies in $C(\tau)$ allow us to detect the precise scale at which topological symmetry breaking occurs, providing a bridge to the structural analysis of defects that follows.}
\section*{Percolation of topological defects}
We still lack any indicator of a structural phase transition, as the observed fraction of negative links does not have an immediate interpretation in terms of site or bond percolation problems \cite{Stauffer2003}. \new{A complementary physical interpretation comes from the heat-like equation $\de_t\psi = -\Slapl\psi$: near the defect-percolation threshold, the diverging correlation length motivates an effective coarse-grained description in which $\Slapl$ behaves as a Laplacian perturbed by localized sink-like terms, i.e. $\Slapl \approx -\Lapl + B$, where $B$ encodes the coarse-grained effect of antagonistic links (see Methods and \gls{si}).} In particular, it can be easily demonstrated that a negative link generates an ``effective'' influence area, affecting the surrounding region. It comes natural to speculate that the inclusion of single isolated negative links can be interpreted as a continuum percolation problem, where the influence area will depend on the specific underlying topology \footnote{We use the term ``topology'' in the graph-theoretic sense, referring to connectivity patterns and frustrated cycles in signed graphs, and not in the sense of topological invariants in condensed-matter physics.}. Although this illustration does not resolve our original problem, it offers a valuable physical interpretation of the phenomena that occur as antagonistic interactions are progressively introduced to the system.

Inspired by the fundamental stabilizers proposed by Kitaev in the realm of topological computation (essential for error detection and correction in the surface code model \cite{Kitaev2006}), we propose and analyze the three fundamental \emph{topological defects}: X defects, that can be mapped to nodes with all negative links, symbolizing disrupted interactions, Z defects, that correspond to frustrated plaquettes, or elementary system cells (see \FigRef{fig:g:percolation_panel}), and the already discussed single defects (S) corresponding to negative links. 
From now on, $p$ corresponds to the relative number of defects included in the system.

We propose to use the components of the eigenvector ($\ket{\lambda_0}$) corresponding to the smallest eigenvalue $\lambda_0$ of $\Slapl$ to define a universal order parameter for the emergence of a topological phase transition. Specifically, we take the signs of $\ket{\lambda_0}$ to identify independent mesoscopic clusters, with the most abundant elements assigned a positive sign. Note that in the limiting case $p=0$, $\ket{\lambda_0}$ is a uniform vector, but as $p$ increases, it develops a mix of positive and negative components, capturing how the system fragments into separate clusters. This transition naturally connects to percolation theory, offering a general framework to analyze it. The lowest eigenmode thus plays a dual role, capturing both structural fragmentation and the long relaxation bottlenecks induced by near-zero eigenvalues.

In particular, $\ket{\lambda_0}$ captures how the system fragments when global translational invariance --and with it, the unique ultraviolet cutoff, $\Lambda$-- is lost. As a result, the ordered (ferromagnetic) phase spanning the entire system, represented by a uniform vector, is no longer a physically viable solution. \FigRef{fig:a:percolation_panel} illustrates different system configurations at criticality, above, and below the expected percolation critical threshold for 2D square lattices, while \FigRef{fig:b:percolation_panel} presents the same for 3D cubic lattices when S defects are introduced. We must make just one final technical remark: since in regular lattices with $p=0$, $\lambda=k^2$, small eigenvalues correspond to large spatial scales in $k-$ space. Thus, numerical precision sets a resolution limit on the maximum system size that can be reliably analyzed for a given topology. This constraint arises from double-precision arithmetic, as discussed in detail in \gls{si}.

\FigRef{fig:c:percolation_panel} shows the percolation phase transition for a 2D square lattice where a double-peak percolation transition emerges. There, the fraction of points belonging to the largest cluster, $P_\infty=n_\infty/N$, is the order parameter, with $N$ the total system size and $n_\infty$ the number of sites belonging to the percolating cluster. The intrinsic fluctuations, $\chi\equiv \sqrt{N} \sigma_{P_\infty}$, diverge at $p_\crit$ as expected for usual second-order phase transitions \cite{LG}. As reported in the inset of \FigRef{fig:c:percolation_panel}, the first peak in the susceptibility, located at $p_\crit\sim0.10(1)$, is fully compatible with the expected threshold for frustrated plaquettes and/or the emergence of the spin glass phase at zero temperature \cite{Queiroz2006}. Further examples considering triangular and hexagonal lattices are reported in \gls{si}, predicting the threshold where spin glass phases emerge \cite{Queiroz2006, miyazaki2013frustration} (we will discuss this important point below). Moreover, \FigRef{fig:e:percolation_panel} shows the percolation phase transition for a 3D cubic lattice, where a double peak in the system susceptibility emerges, with $p_\crit=0.22(1)$ \cite{Hartmann1999} (4D analyses are reported in \gls{si}). The cluster-size distribution (see \FigRef{fig:d:percolation_panel} and \FigRef{fig:f:percolation_panel}) confirms that the phase transition effectively belongs to the ordinary percolation universality class, as $P(S) \sim S^{-\tau}$, with $\tau\approx2.18(1)$ for the 3D case. The case of 2D lattices is more subtle, as we observe an exponent $\tau\approx1.75(1)$, still compatible with previous results in small-to-mid systems in continuum percolation problems, where logarithmic corrections may play a key role in the convergence of the cluster size distribution for low dimensions \cite{JSTAT_Poisson}.

The double peak in the susceptibility represents a delicate point that must be discussed in terms of the geometrical properties of percolation in Euclidean spaces. Specifically, in three dimensions, two spanning clusters can geometrically coexist due to the natural emergence of lower-dimensional percolation structures. This is a natural consequence of the Alexander-Orbach conjecture \cite{AO}: as critical percolation clusters present a universal spectral dimension $d_S=\nicefrac{4}{3}$ they can only coexist in dimension $d\geq3$. In fact, we observe how the double-peak structure represents a finite-size effect in 2D lattices, which disappears in the infinite-size limit. Instead, in 3D, this behavior is consistent with survival in the thermodynamic limit (see \gls{si}, and being compatible with observations of coexistence of spin glass and ferromagnetic order in 3D systems \cite{Beath2007}).

We report in \FigRef{fig:h:percolation_panel} and \FigRef{fig:i:percolation_panel} the phase transition when, respectively, Z or X defects are included in a 2D lattice. In particular, for both cases, the percolation threshold is equivalent to those of S defects, as they exhibit a percolation phase transition for the same values of $p_\crit$ (see also \gls{si} for further analyses on different lattices). The case of X defects generates an entirely new situation. In fact, in the squared 2D lattice, the double-peak structure does not vanish in the thermodynamic limit, as the second peak is directly related to the site-percolation problem. Note that X defects in the low-density regime are not expected to produce any closed loop, leading to no local frustration of interactions (the system is, then, effectively balanced, i.e., the fundamental eigenvalue is equal to zero). \new{Balance, however, does not imply trivial dynamics, as mesoscale fragmentation and competing basins can still emerge.} This leads to new fundamental dynamic implications that are explored below when X defects are considered together with a dynamic process.

\begin{figure*}[hbtp]
    \centering
    \includegraphics[width=\textwidth]{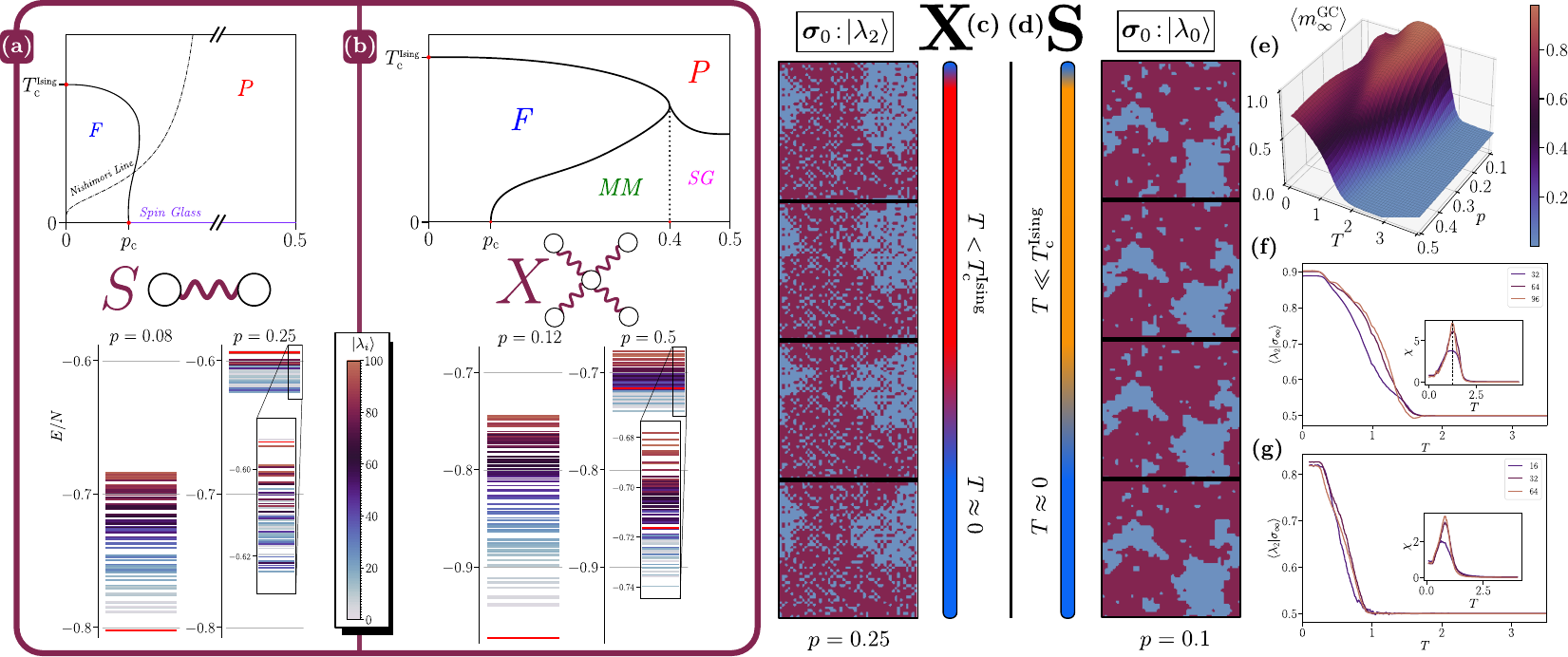}
    \caption{\textbf{Ising dynamics. (a) } Qualitative \(T-p\) phase diagram for the 2D squared lattice with S defects. The lower part shows the energy levels corresponding to the first 100 eigenstates of $\Slapl$ (see colorbar), for two values of \(p\), respectively, in the ferromagnetic phase and the spin-glass one. The lowest eigenstate of $\Slapl$ has been marked in red. Note that, inside the spin-glass phase, all eigenstates come closer in energy, and their ordering is lost at the 'onset of degeneracy' at $p_\crit$. \textbf{(b)} Qualitative \(T-p\) phase diagram for X defects computed with the \gls{sl} eigenstates through numerical simulations in a 2D square lattice of size \(N=64^2\). The breakup of the ferroelectric phase translates into the emergence of a metastable one where the \gls{sl} eigenvectors now define multiple degenerate stable states characterized by local order up to the paraelectric/ferroelectric phase transition. At a certain $p$, stable minima becomes unstable, leading to a spin-glass phase. The lower part shows the energy levels corresponding to the first 100 eigenstates. In this case, the energy degeneracy is not observed before \(p=0.4\), where the spin-glass phase emerges. \textbf{(c)} Annealing and sudden quenching initializing the system in a low eigenmode \(\ket{\lambda_2}\) spin configuration on a 2D squared lattice with \(p=0.25\) fraction of X defects. We heat the system up to \(T\approx1.5\), i.e., smaller but closer to the critical Ising temperature. After relaxation, the system is suddenly quenched, observing that, up to border effects, it does not escape from the local minima associated to \(\sign(\ket{\lambda_2})\). \textbf{(d)} Same as the previous case but using S defects with the lowest \gls{sl} eigenmode at the $p_\crit$. In this case, the system is locally stable but starts the attractor surfing at much lower temperatures. \textbf{(e)} Local magnetization order parameter, $m^{\rm GC}$, as a function of the temperature and the fraction of X defects for $N=64\times64$. \textbf{(f)}- \textbf{(g)} Spin overlap, $\braket{\lambda_i}{\spin_\infty}$, as a function of the temperature for different lattice sizes for (f) X defects, using $\ket{\lambda_2}$, and (g) S defects, using $\ket{\lambda_2}$. The critical temperature is indicated in the insets as a vertical dashed line. Note how $T_\crit$ is significantly greater for the case of X defects, while, in contrast, the initial configuration loses its stability at much lower temperatures for S defects. All curves have been averaged over \(n_{\rm avg} = 1000\) independent realizations.}
    \label{fig:rbim_pd}
\end{figure*}

\section*{Exploring the dynamical arena of antagonistic non-ergodic systems}

In frustrated systems, the emergence of non-ergodic dynamics manifests in the existence of multiple stationary states with disjoint basins of attraction \cite{parisi2023nobel}. Hence, predicting stationary states in such systems at low temperatures represents a significant challenge due to the intrinsic geometrical complexity of the problem. We state the following conjecture: for a vanishing thermal noise, i.e., at zero temperature, these effects are fully described by the topological symmetry-breaking mechanism reported in the previous section. Their emergence will thus be strictly linked to the percolating clusters defining the structural shape of the system. Specifically, we have analyzed the dynamical implications of the different topological defects simulating the \gls{ea} Hamiltonian (see Methods),
\begin{equation}
    \ham=-\underset{\langle i,j\rangle}{\sum}\jij_{ij} \spin_i\spin_j,
    \label{eq:ising_sg_hamiltonian}
\end{equation}
where \(\spin_i\) is the spin in the \(i^{\rm th}\) site, with \(\jij_{ij} = \jij \adj_{ij}\), being $A$ the signed adjacency matrix, and \(\jij\) the global coupling strength.

\FigRef{fig:rbim_pd}(a) and (b) show the temperature phase diagram of the system as a function of the fraction of negative links, $p$, for S and X defects on 2D square lattices, respectively. In particular, S defects correspond to the well-known \gls{rbim} or \(\pm \jij\)-model \cite{toulouse1977theory, edwards1975theory} (see Methods). Results are achieved through extensive Monte Carlo simulations of spin-spin dynamics, running \eqref{eq:ising_sg_hamiltonian} on top of the signed structure. The inspection of the phase diagram confirms the well-known collective dynamical regimes for S defects: ferromagnetic and spin glass phases. Now, we characterize the eigenstates of the system using the corresponding signed eigenvectors of $\Slapl$ to define topologically-based up and down spin islands. Then, we calculate the energy of the system at $T=0$ through the \gls{ea} Hamiltonian as defined in \eqref{eq:ising_sg_hamiltonian}. This confirms the degeneracy of energetic states at the critical fraction of negative links, $p_\crit$, for both S and X defects. \FigRef{fig:rbim_pd}(a) and (b) report how the ground state energy gap disappears, leading to the degeneracy and, finally, inversion of the energetic levels of the system. Note that, for X defects, the diagram shows, however, distinct bifurcation lines separating new phases that are still unknown to the best of our knowledge. In particular, for low temperatures, increasing the fraction $p$, a first transition separates the ferromagnetic phase from a metastable one, where the system can get trapped into local minima up to the Ising critical temperature, $T_\crit$ (see \FigRef{fig:rbim_pd}(c)). A second transition at higher $p$ leads to the spin glass phase, where the resilient and locally stable spin islands become unstable when thermal noise is applied. Notably, this new intermediate phase between the ferromagnetic and spin glass phases, the MM phase, exhibits multiple coexisting local islands of mesoscopic magnetic order. Note that, unlike the chaotic attractor ``surfing'' seen in spin glass phases at low temperatures, the MM phase allows access to highly diverse noise-resilient functional configurations.

Note that, for S defects, the usually reported intriguing re-entrance of the ferroelectric phase to slightly higher values of $p$ when the thermal noise increases \cite{Thomas2011} can be ascribed to the breaking of small frustrated clusters due to noise effects. Specifically, thermal noise induces random flips with a single-spin interaction radius that is explicitly smaller than the size of the clustered defects. This effect first breaks small spin-clusters, reducing domain wall effects and making more robust the giant ferroelectric cluster. 

\FigRef{fig:rbim_pd}(c) and (d) illustrate the stability of different eigenstates of $\Slapl$ as the system is heated up to different temperatures. We emphasize that, for X defects, the system can be heated up to the Ising critical temperature, where it is reset by noise, and allow for choosing other possible local minima. The same occurs at a much lower temperature, $T_{\rm SG}(N)\ll T_\crit^{\rm Ising}$, for S defects.  In the latter case, typical of spin glass behavior, the minima are locally stable, but the system can escape under low noise levels, exploring the phase space (see \gls{si}). We hypothesize that the spin-glass tree of states \cite{parisi2023nobel} is expected to be a superposition of the \(N\) eigenvectors of the \gls{sl}, or \new{or configurations that can approximate \emph{pure states}}, that represent the original set of local minima.

It comes natural to introduce topologically-based local order parameters as a function of $T$ that effectively capture the emergent mesoscopic order when spontaneous symmetry breaking forms magnetized spin islands (see Methods).  \FigRef{fig:rbim_pd}(e) shows the magnetization of the giant cluster of $\ket{\lambda_2}$ as initial state, as a function of $p$ and $T$ for the case of X defects. \new{We examine a complementary mode to $\ket{\lambda_0}$ to probe spectral multiplicity and the presence of competing metastable structures.} Hence, \FigRef{fig:rbim_pd}(f) and (g) show the spin overlap $\braket{\lambda_i}{\spin_\infty}$ between the initial state and the long-time dynamics when the system has thermalized (see Methods). In particular, for the case of X defects, this evidences how when the system is initialized at $\ket{\lambda_2}$, it undergoes a second-order phase transition with a critical temperature a little below the Ising critical one, $T_\crit\simeq2.27$. This provides direct evidence of the stability of the corresponding eigenstate or initial configuration of the system. By contrast, \FigRef{fig:rbim_pd}(g) shows the narrow temperature range in which the selected starting eigenstates for S defects lose their stability, causing the system to start surfing across other attraction basis, i.e., pure states, or combinations thereof.

\begin{figure}[hbtp]
    \centering
    \includegraphics[width=1.0\columnwidth]{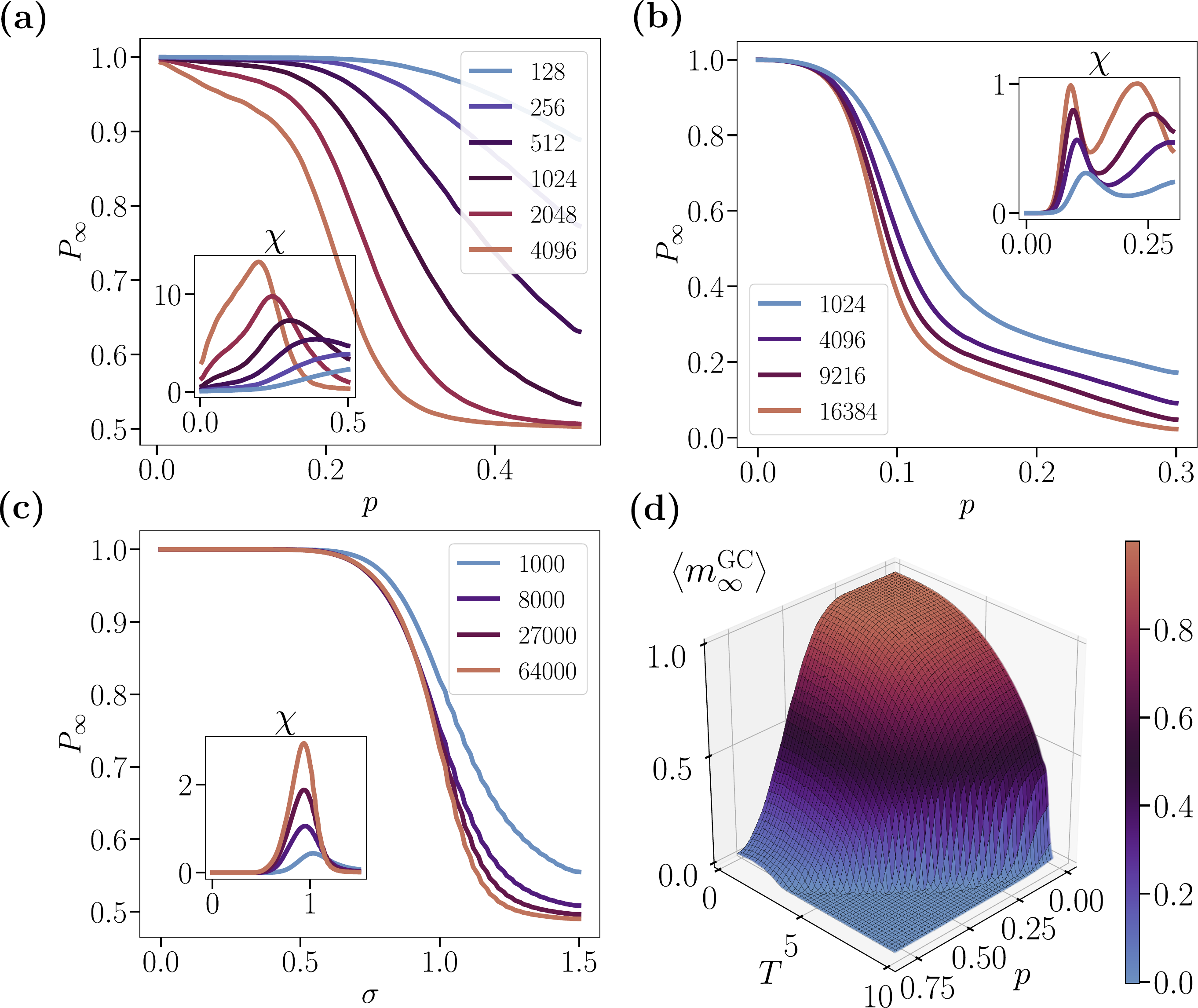}
    \caption{\textbf{Heterogeneous architectures}. Order parameter, $P_{\infty}$ versus the fraction of negative links, $p$ for a: \textbf{(a)} \gls{er} network with $\langle \kappa \rangle=10$, \textbf{(b)} Small World network from a 2D lattice (rewiring probability \(p_\rew=0.05\)) and \textbf{(c)} a diluted 3D lattice ($p_\dil=0.25$) with normally weighted edges \(p\sim\mathcal{N}(1, \sigma)\). Inset: Variance of the order parameter, $\chi$, as a function of the fraction of negative links, $p$. \textbf{(d)} \(T-p\) Phase diagram of the \gls{er} network with \new{a fraction $p$ of} X defects \new{measured by} the local Ising order parameter $m^{\rm GC}_\infty$, \new{and using} \(\sign(\ket{\lambda_2})\).}
    \label{fig:phase_trans_hetero}
\end{figure}

\begin{figure*}[hbtp]
    \centering
    \includegraphics[width=.9\linewidth]{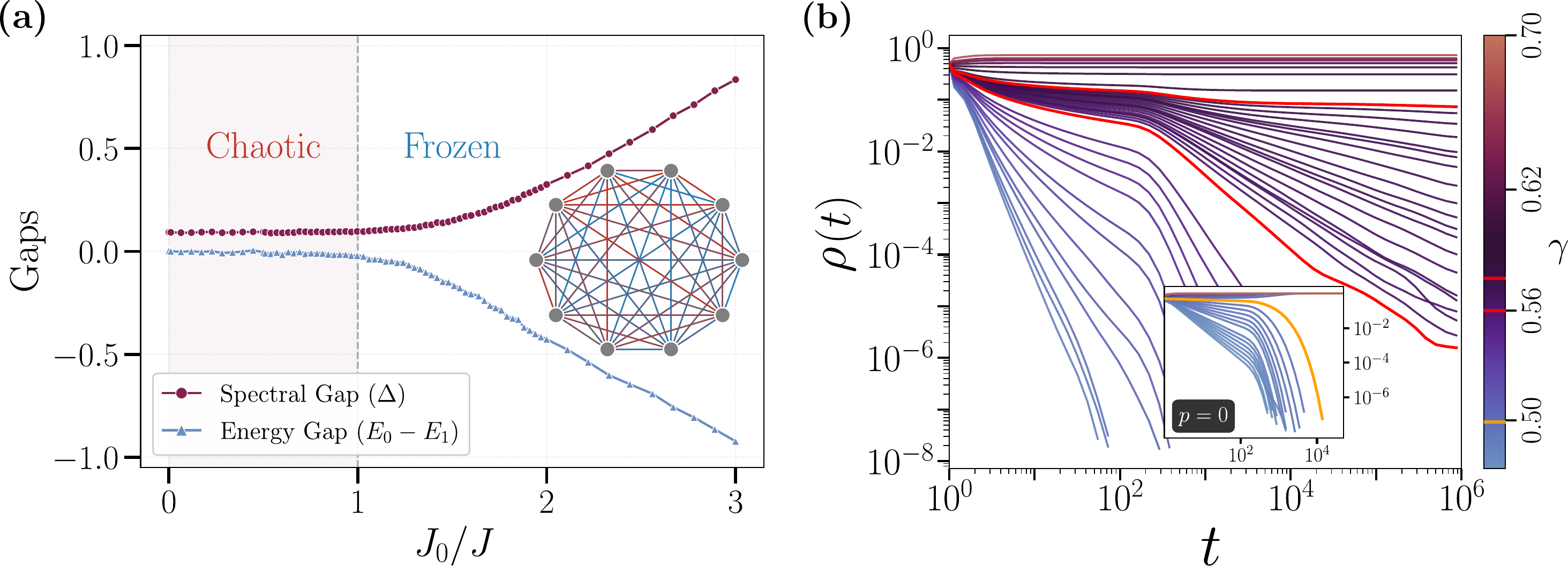}
    \new{\caption{\textbf{(a)} Spectral gap ($\Delta$) and Energy Gap ($E_0-E_1$) versus $J_0/J$ for a fully connected network with symmetric Gaussian interactions, highlighting the closure of the gap at the transition point. \textbf{(b)} Activity decay for the antagonistic contact process with a fraction of inhibitory links, $p=0.15$. Note the slow decay region indicative of Griffiths phases that emerge beyond $p>0.1$. This confirms that the previously identified topological phase transition is consistent with the interpretation that it governs the divergence of dynamical timescales. This dynamical slowing down is analytically linked to the proliferation of vanishingly small eigenvalues in the Signed Laplacian spectrum, which emerge as topological defects disrupt the spectral gap. Thus, the topological fragmentation of the network provides a structural encoding of the broad distribution of relaxation timescales observed in the dynamics. Inset: Temporal decay for a 2D lattice with no inhibitory links ($p=0$), showing standard bistable behavior. Parameters: L=96.}}
    \label{fig:Sompo}
\end{figure*}

\section*{Topological defects in heterogeneous architectures}
To establish whether the \gls{sl} is able to predict a percolative phase transition in heterogeneous systems, we have analyzed different dilute models and network structures by introducing a random fraction of S defects. We highlight that this approach generalizes the detection of frustrated dynamical phases to complex networks and generic disordered systems.

Specifically, the results for \gls{er} and \gls{sw} networks are reported in \FigRef{fig:phase_trans_hetero}(a) and (b). Additionally, we have analyzed a particularly complex case with significant physical relevance: the 3D diluted cubic lattice, which has originally inspired dilute Ising models \cite{Harris1974}. Now, a fraction $f$ of the existent links are removed, remaining below the edge-percolation threshold to ensure the system is connected. We then assign positive and negative weights to the remaining interactions, randomly drawn from a Gaussian distribution with mean $\mu$ and variance $\sigma$. As illustrated in \FigRef{fig:phase_trans_hetero}(c), we also predict the emergence of the frustrated phase as a function of $\sigma$. Finally, \FigRef{fig:phase_trans_hetero}(d) shows the \new{local Ising order parameter} for the case of \new{ including a fraction $p$ of X defects in an} \gls{er} network.

\new{\section{Implication for E/I models of neural networks}
Antagonistic interactions, central to the balance between excitation and inhibition in neural circuits, have long been a cornerstone of theoretical neuroscience. Moving beyond classical spin-lattice approaches that typically give rise to short-range spin glass models, we consider alternative formulations that provide a basis for more refined descriptions of neural computation. The first is the Sompolinsky--Crisanti--Sommers (SCS) model~\cite{Sompolinsky1988}, which describes the dynamics of large recurrent networks with mixed excitatory and inhibitory synaptic interactions. This framework offers a minimal theoretical explanation for the intrinsically irregular, yet collectively organized population dynamics observed in cortical circuits~\cite{Litwin2012}. The model can be formulated in terms of continuous activity variables $x_i(t)$ that interact via quenched couplings $J_{ij}$ and are transformed through a nonlinear transfer function $\phi(\cdot)$; a complete treatment is provided in the Methods. Treating the connectivity as quenched disorder and applying dynamical mean-field theory~\cite{Stern2014}, the model predicts a transition from a stable fixed-point regime to high-dimensional chaotic activity when the interaction strength exceeds a critical threshold. While the fully asymmetric case yields high-dimensional chaos, the onset of this instability is universally marked by the destabilization of the trivial fixed point (determined by the symmetric part of the interaction spectrum).   As reported in \FigRef{fig:Sompo}5(a), we observe that this critical point coincides with the closure of the spectral gap $\Delta$ of the underlying Signed Laplacian. Furthermore, the energy gap $E_0 - E_1$ between the ground state and the first excited state of the effective energy landscape vanishes at the same threshold. This supports the view that the loss of stability in continuous rate models—regardless of whether the resulting phase is chaotic or glassy—is rooted in the same topological symmetry-breaking mechanism identified in discrete spin systems.

The second model we consider is a simple two-state (binary) neural network model \cite{Larremore,Jensen}, in which each unit is either active ($s_i(t)=1$) or inactive ($s_i(t)=0$) at time $t$. The state of unit $i$ evolves according to the weighted activity of its $k$ neighbors,
$\mathcal{P}_i = f\!\left(\Lambda_i = \frac{\gamma}{k} \sum_{j} \omega_{ij} s_j(t) \right)$, where $\omega_{ij}=\pm 1$ encodes excitatory and inhibitory interactions and $\gamma$ controls the interaction strength. 
To connect with the symmetric interaction framework used in this work, we consider a variant in which excitation and inhibition are assigned to \emph{links} rather than nodes, so that $\omega_{ij}=\omega_{ji}$ and a fraction $\alpha$ of links are inhibitory. 
This link-based formulation preserves the global E/I balance while ensuring a symmetric coupling matrix, allowing us to isolate the collective effects of antagonistic interactions; full details are provided in the Methods section. \FigRef{fig:Sompo}5(b) shows the temporal decay of the network activity from a fully active state. For low fractions of inhibition ($p < 0.1$), the system exhibits a single critical point with rapid relaxation or simple bistability (see Inset). 
However, beyond the critical threshold $p \approx 0.1$, we observe the emergence of 'Griffiths phases'--regions characterized by an anomalously slow, power-law decay of activity. This phenomenon is not merely a dynamical artifact but a direct consequence of the spectral properties of the Signed Laplacian: as topological defects percolate, they induce a proliferation of degenerate eigenvalues. 
Since relaxation timescales are inversely proportional to these eigenvalues ($\tau \sim 1/\lambda$), the topological fragmentation of the network structurally guarantees the emergence of the diverse, long-lasting timescales often associated with criticality in neural circuits.}

\section{Discussion and conclusions}


\new{We provide a new perspective on a decades-long controversy} surrounding the nature of spin glasses. We have explicitly shown how non-ergodic phases originated by competing interactions, such as the emergence of the spin glass phase at zero temperature, lie in a pure topological symmetry-breaking mechanism that goes beyond the considered dynamics. This mechanism exhibits a clear algebraic signature in the emergence of degeneracy of the ground state of the \gls{sl} operator. Topological defects generate natural barriers in the discrete lattice that lead to the fragmentation of the system in mesoscopic regions already at low values of $p$. Naturally, such a mechanism strongly depends on the discrete nature of the underlying lattice, leading to different percolation thresholds even if the spatial dimension of the system remains unchanged. \new{While the Signed Laplacian does not fully determine nonlinear dynamics, it provides a minimal structural backbone that constrains the landscape on which these dynamics unfold.}  Hence, the \gls{sl} framework offers a natural approach for detecting spin glass phases in any homogeneous or heterogeneous structure. 
This has profound implications for considering future dynamical \gls{rg} analysis as it directly acts on the Green function of the system by adding only a new term correcting $\Lapl$, which hampers the topological properties of the embedding space on top of which any field-theoretical approach is based.

It is precisely the local nature of particular defects that make different non-ergodic systems strongly unstable. For example, in the case of S defects that lack local stability, frustrated loops of arbitrary size can be generated by thermal noise, leading to large avalanches that cause the system to escape from a stable minimum, allowing it to explore large parts of the available space. In other words, frustrated loops are expected to induce noise effects, leading the system to a sempiternal out-of-equilibrium condition that permits it to make attractor surfing easier. This last case corresponds to the Z and S topological defects, making spin glasses elusive when searching for stable configurations. A different picture can be drawn for X defects: the system shows multiple minima now, but high energy barriers allow the system to remain trapped in each one, making it predictable. However, we want to emphasize that all defects generally lead systems to ergodicity breaking, where a single stable state no longer exists.

The percolative analysis presented here also permits an immediate interpretation of the long-standing debate of the upper critical dimension of spin glass systems \cite{parisi2023nobel}. That is, on determining the dimensionality $d_u$ above which the critical behavior of spin glasses is described by mean-field theory. Suppose one assumes that the dimension of the percolative giant cluster determines the effective dimension where the spin system operates. In that case, $d=6$ is the upper dimension for the isotropic percolation universality class, which presents a fractal dimension $d_f=4$, which is precisely the upper dimension for the Ising universality class. Hence, specific spatial dimensions $d=4$ and $d=5$ are known to exhibit a fractal cluster with $d_f\approx3.04$ and $d_f\approx3.52$, respectively \cite{Jan1985}, making invalid the mean-field solution for Ising dynamics when competing interactions are considered. 

\new{ 
Percolation phenomena in correlated and higher-order networked systems have also been investigated in other contexts, such as multiplex and hypergraph models or reaction networks \cite{Sun2021,Nicolaou2023}. While these studies rely on different mechanisms, they are consistent with the broader view that structural organization can influence collective transitions and may motivate future extensions of the present framework.}

\new{One might ask whether other spectral metrics, such as the standard Laplacian or measures of diffusion capacity, could capture these phenomena. Standard graph operators, however, typically treat links as purely conductive channels (positive weights), thereby ignoring the crucial algebraic structure of antagonistic interactions. While metrics like the Diffusion Capacity describe transport efficiency, they fail to detect the frustration arising from negative cycles. The Signed Laplacian is unique in this regard: it naturally incorporates the sign of the interaction into the diffusion process, ensuring that the spectral gap $\Delta$ directly measures the structural balance of the network. Consequently, the Signed Laplacian provides a direct and physically interpretable link between the algebraic structure of antagonistic interactions and the effective energy landscape governing relaxation dynamics.}

Finally, our results significantly progress in defining natural stable states, determined by the \gls{sl} operator, where frustrated systems can exhibit stable minima. Moreover, we emphasize that this step helps to reduce an NP problem --solved by hit-or-miss methods-- to a combinatorial one of assembling system eigenstates as fundamental building blocks. 

Our framework also opens a route to extend this approach to learning algorithms, helping to develop a common mathematical framework that better characterizes non-ergodic phases that facilitate engineered computation depending on specific microscopic details.

\section{Acknowledgments}
P.V. acknowledges the Spanish Ministry of Research and Innovation and Agencia Estatal de Investigaci\'on (AEI), MICIN/AEI/10.13039/501100011033, for financial support through Project PID2023-149174NB-I00, funded also by European Regional Development Funds, as well as Ref. PID2020-113681GB-I00. We also thank G. Cimini, M.A. Mu\~noz, P. Caputo, and C. Bordenave for their useful suggestions and comments.

\section{Methods}

\paragraph{The discrete signed Laplacian.} The traditional definition of the Laplacian operator, represented by \(\lapl = \degm - \adj\), presents serious issues when dealing with signed networks as it is no longer semi-definite positive. This is due to the fact that diagonal terms may take null or even negative values and $\degm^{-1/2}$ and, therefore, $\lapl_{\rm RW}$ is no longer defined.  To tackle this delicate point, a recent proposal has mathematically introduced the signed version of the Laplacian operator. It is grounded in sound physical principles, being capable of accurately capturing the complex and intricate diffusion dynamics of any network \cite{kunegis2010spectral}. In particular, the \gls{sl} of a signed architecture can be defined as,
\begin{equation}
    \Slapl = \sdegm - \adj,
    \label{SignedL}
\end{equation}
where \(\sdegm\) represents the unsigned degree matrix, i.e., the diagonal matrix given by $\sdegm_{ii} = \sum\nolimits_j\abs{\adj_{ij}}$, with $\adj$ the signed adjacency matrix.

From \eqref{SignedL}, it follows that the \gls{sl} can be rewritten as the usual Laplacian $\Slapl$ of the unsigned version of the network minus two times the adjacency matrix of the negative links $\adj^-$, namely,
    \begin{equation}
        \Slapl = \lapl - 2\adj^-,
    \end{equation}
which, in the continuum approximation, can be seen as
    \begin{equation}
        \Slapl \equiv -\Lapl + B.
        \label{expansion}
    \end{equation}
where $B$ represents a generic operator affecting the usual second derivative term, tantamount to sink effects.

\paragraph{Properties of  the signed Laplacian.}

The \gls{sl} operator preserves relevant properties, namely:\par
\begin{enumerate}[label=(\roman*)]
    \item The \gls{sl} matrix \(\Slapl\) is positive-semidefinite for any graph.
    \item $\Slapl$ becomes strictly positive-definite for the unbalanced graphs (that is, for connected unbalanced networks, there is only one eigenvalue strictly equal to zero, $\lambda_{\min}=0$, in the total absence of negative loops).
    \item $\lambda_{\min}$ measures the level of balance (frustration) of the network \cite{kunegis2010spectral}.
\end{enumerate}

\paragraph{SL as diffusion dynamics in the presence of sinks.}
In particular, \eqref{expansion} can be reduced, when considering only one negative link, to the problem of point-like particles diffusing in a uniform medium and being absorbed by a single spherical sink \cite{felderhof1982diffusion} (see further details in \gls{si}), with \(m = m(\vec{x}, t)\) the magnetization field,
\begin{equation}
    \pdv{m}{t} = D_0\Lapl m -\lambda(\vec{x})m + s(\vec{x}, t),
    \label{sinks}
\end{equation}
where $D_0$ stands for the diffusion coefficient, $\lambda(\vec{x})$ is the local absorption coefficient and \(s(\vec{x}, t)\) the source density.
\paragraph{Field-theoretical implications.}
It is well-known that the Green function for the Gaussian model follows the expression,
\begin{equation}
    G(\textbf{k},\omega)=[-i\omega+D_0k^2]^{-1}
\end{equation}
Through the extension of \eqref{sinks} to many sinks \cite{felderhof1982diffusion} one can find the following Green function,
\begin{equation}
    G(\textbf{k},\omega)=[-i\omega+D_0k^2+nt(\textbf{k},\textbf{k},\omega)]^{-1}
\end{equation}
where the last term represents the interaction with sinks in the system. This opens the door to the rigorous formulation of a field-theoretical framework for spin glasses and other frustrated dynamics as, in principle, the signed interactions add an extra term to the Green function, which is the basis of diagrammatic expansions of the correlation function.

\paragraph{Canonical formulation of signed interactions.} Once we can consider a semidefinite operator for signed networks, $\Slapl$, it is possible, in terms of the network propagator, $\hat K=e^{-\tau\Slapl}$, to generalize the ensemble of accessible information diffusion states \cite{dedomenico2016spectral,LRG,InfoCore,NatRevPhys}, namely,
\begin{equation}
 \dm(\tau)= \frac{e^{-\tau\Slapl}}{Z},
 \label{EvolMat2}
\end{equation}
where $\rho(\tau)$ is tantamount to the canonical density operator in statistical physics (or to the functional over fields configurations) \cite{Binney, Pathria}, and $Z(\tau)=\sum_i^N e^{-\tau\lambda_i}$, being $\lambda_i$ the set of system's eigenvalues. It is possible to, therefore, define the network entropy \cite{dedomenico2016spectral} through the relation
\begin{multline}
 S[\dm(\tau)]=-\Tr\left[\dm(\tau) \log \dm(\tau)\right] = \tau \avg{\lambda}_{\tau} + \log Z (\tau)
 \end{multline}
being $\langle \hat O \rangle_{\tau}=\Tr[ \dm (\tau) \hat O ]$. Immediately, it is possible to define the heat capacity of any signed network as \cite{InfoCore},
\begin{equation}
 C=-\dv{S}{(\log \tau)}=-\tau^2\dv{\avg{\lambda}_{\tau}}{\tau}
 \label{SHeat}
\end{equation}
\eqref{SHeat} describes the heat capacity of the network, which can detect transition points that correspond to the intrinsic diffusion scales characteristic of the network. When condition \(\dv*{C}{\tau}|_{\tau^{*}}=0\) is met, it defines $\tau^*$ and reveals pronounced peaks that reflect a significant deceleration in information diffusion. If \(C\) also shows a well-defined plateau, it represents direct evidence of a scale-invariant architecture \cite{LRG,JSTAT,Poggialini} (such as, for example, 2D lattices). In this case, a peak at short times, in turn, independent of system size, captures the shortest resolvable interaction scale (by analogy with an ultraviolet cutoff), denoted $\Lambda$ in lattice settings \cite{TBM}. This cutoff represents a minimum scale that prevents divergence in usual \gls{rg} calculations \cite{Kardar} and indicates that the system exhibits some kind of translational invariance if the subsequent plateau is present \cite{Poggialini}. 

\paragraph{Signed Ising dynamics.} The archetypal spin glass model is defined as the celebrated \gls{ea} model \cite{edwards1975theory}, with the corresponding Hamiltonian
\begin{equation}
    \ham = -\sum_{(i,j)}\jij_{ij}\spin_i \spin_j,
\end{equation}
where \(\spin_i=\pm1\) are Ising spins, \((i,j)\) are the nearest-neighbors' bonds of the network and $\jij_{ij}=\jij\adj_{ij}$, with $\adj_{ij}$ the weighted and signed adjacency matrix. This represents the paradigmatic model for \emph{disordered} magnets with short-range spin-spin interactions, including multiple negative loops.

Note that the previous equation can be safely rewritten as \cite{Binney},
\begin{equation}
   \ham = -\jij\sum_{(i,j)}\adj_{ij}\spin_i \spin_j+\jij\sum_{(i,j)}\ndegree_i\delta_{ij}\spin_i\spin_j, 
\end{equation}
where $\ndegree_i$ is the node connectivity. This automatically leads to the natural definition of the \gls{sl} as written in \eqref{SignedL}.

We have set, for simplicity, \(\jij = 1\). For the case of S defects, the \(ij^{\rm th}\) value is a random variable extracted from the probability density function $\prob(\jij_{ij}) = (1-p)\ddelta(\jij_{ij}-\jij) + p\ddelta(\jij_{ij}+\jij)$, which correspond to the \gls{rbim}. Regarding X defects, the parameter \(p\) denotes the probability that a site or node is selected for flipping, which entails that all edges incident on that node are assigned a negative weight. 
In contrast, for Z defects, the parameter \(p\) represents the fraction of unit cells, i.e., closed loops within the lattice in which every edge is flipped.

The global normalized magnetization of the system is naturally defined as,
\begin{equation}
   m=\frac{1}{N}\sum_{i=1}^N\spin_i
\end{equation}
where $N$ corresponds to the system size and $s_i$ is the state of the Ising spins at site $i$.

Instead, we can measure the local order of the system through different observables when using some selected \gls{sl} eigenstates, $\ket{\lambda_i}$ as initial states. To perform this, the system is prepared in a selected 'pure' eigenstate, where negative and positive signs are associated to up and down states, respectively.

One possibility corresponds to the so-called spin overlap, i.e. the superposition between the initial state and the dynamical evolution of the system at each time $t$, as follows,
 \begin{equation}
    \braket{\lambda_i}{\vec\spin(t)} = \frac{1}{N}\sum_{k=1}^N\delta(\sign(\lambda_i^{(k)})-\spin_k(t))
\end{equation}
where $\lambda_i^{(k)}$ is the k-th component of the i-th eigenmode $\ket{\lambda_i}$, with $\vec{\spin}$ the vector of magnetizations at time $t$.

Alternatively, one can measure the magnetization of the giant cluster of the system in its initial state. Then, the local order parameter runs over the $n_\infty$ spins of the giant component of this specific state as indicated by $\ket{\lambda_i}$,
\begin{equation}
   m^{(i)}_\infty(t)=\frac{1}{n^{(i)}_\infty}\stackrel[j=1]{n^{(i)}_{\infty}}{\sum}\spin_{j}(t)
    \label{eq:magn_inf}
\end{equation}
\paragraph{\textbf{RBIM vs field theory at $T\to 0$.}}

In order to tighten the relation between the fundamental state of the \gls{sl} and the ground state of the Hamiltonian \eqref{eq:ising_sg_hamiltonian} with the random interaction matrix \(\jij_{ij}\), it is convenient to introduce the field theoretical formulation of the Ising model given by the Hubbard-Stratonovich transformation \cite{Amit} for which
\begin{equation}
    \label{H-S-transform}
    \begin{aligned}
        \partf(\jij) &= \sum_{\vec\spin} e^{-\beta \ham[\vec\spin]} \\
    &\propto \int \fundiff\dfield\,\exp\bigg\{
    -\beta\sum_{(i,j)}\jij_{ij}\dfield_i\dfield_j \\
    &\quad+ \sum_i \ln\Big[\cosh\Big(2\beta\sum_j \jij_{ij}\dfield_j\Big)\Big]
    \bigg\},
    \end{aligned}
\end{equation}
where $\beta=1/T$ and $\fundiff\dfield=\prod_i d\dfield_i$.
For $\beta\to +\infty$ we can apply the saddle-point method leading to the self-consistent equations for the ground state:
\[\sum_{i}\jij_{ki}\left(\dfield_i-\sign\left[\sum_l \adj_{il}\dfield_l\right]\right)=0,\]
for $k=1,2, ...., N$, where $\adj_{ij} = \jij_{ij} / \jij$ is the adjacency matrix of the network of interactions. The self-consistent solution of the above equation is
\begin{equation}
    \dfield_i=\sign\left[\sum_l \adj_{il}\dfield_l\right].
    \label{ground-state}    
\end{equation}
Note that, as expected, for the case $p=0$ this gives two specular ferromagnetic solutions either $\dfield_i=+1$ or $\dfield=-1$ for all $i$. The symmetry can be broken either by a small and uniform external magnetic field or by suitable boundary conditions.

At small $p\ne 0$ clearly two ferromagnetic ground states related by the same global up-down symmetry are still present, but they are no more homogeneous due to the inhomogeneous pair interactions. When instead $p$ becomes large enough, i.e. at the $T=0$ critical value $p_c$, the ferromagnetic ground state is no more unique, but many different and degenerate ground states are solutions of \eqref{ground-state} and a spin glass transition appears. 

Let us now see the relation between the solutions of \eqref{ground-state} and the eigenvector(s) related to the least eigenvalue of $\Slapl$.
As aforementioned, it (they) satisfies the following equation
\[\sum_{i=1}^N\Slapl_{ij}\cfield_j=\lambda_0\cfield_i\,,\]
which can be rewritten also as
\begin{equation}
    \sum_{i=1}^N \adj_{ij}\cfield_j=(\ndegree_i - \lambda_0)\cfield_i\,,
    \label{sign-Laplace}
\end{equation}
where $\ndegree_i=\sum_j \abs{\adj_{ij}}$ is the coordination number of site $i$ ($\ndegree_i=2d$ for a $d-$dimensional square lattice). 
As shown in \cite{Liu2015}, $\lambda_0\le \min_{i=1}^N \ndegree_i$. This implies that 
\begin{equation}
\sign[\cfield_i]=\sign\left[\sum_l \adj_{il}\cfield_l\right],
\label{sgn-psi}
\end{equation}
for all connected networks, including regular lattices. \eqref{sgn-psi} is formally the same of \eqref{ground-state} with the fundamental difference that the solution of the former has to be found on the $N-$dimensional spherical surface of radius $\sqrt{N}$ while for the latter has to found on the vertices of the $N-$dimensional cube defined by $\dfield_i=\pm 1$ for each $i=1,2,...,N$ (note however that in both cases $\abs{\vec\psi}=\abs{\vec\phi}=\sqrt{N}$. This implies that $\sign[\cfield_i]$ can differ from $\sign[\dfield_i]$ in some node in particular where there are large fluctuations in the values of $\psi$ between neighboring nodes.
Anyway, being $\vec\phi$ and $\vec\psi$ characterized by the same normalization condition, we expect that  the binarized vector $\vec\cfield_{\rm bin}$ of components $\cfield_i/\abs{\cfield_i}$ shows a similar behavior with respect the binary vector $\vec\phi$ on a suitable coarse-grained scale as large as the frustration walls. 

\new{
\subsection*{Sompolinsky--Crisanti--Sommers (SCS) rate model}

We consider a recurrent network of $N$ units with continuous activity variables $x_i(t)$, generated from effective inputs $h_i(t)$ through a nonlinear transfer function,
\begin{equation}
x_i(t) = \phi\!\left(g\,h_i(t)\right),
\end{equation}
where $g$ is a gain parameter and $\phi(x)=\tanh(x)$ unless stated otherwise. The inputs evolve according to first-order relaxation dynamics,
\begin{equation}
\frac{\mathrm{d} h_i(t)}{\mathrm{d}t} = - h_i(t) + \sum_{j=1}^{N} J_{ij}\, x_j(t),
\label{eq:SCS_dynamics_methods}
\end{equation}
where $J_{ij}$ are quenched synaptic couplings drawn as
\begin{equation}
J_{ij} \sim \mathcal{N}\!\left(0,\,\frac{g^2}{N}\right), \qquad J_{ii}=0.
\end{equation}
In the limit $N\to\infty$, the recurrent input to each neuron becomes an effective Gaussian process, and dynamical mean-field theory (DMFT) yields a self-consistent equation for the autocorrelation $C_x(t,t')=\langle x(t)x(t')\rangle$. This theory predicts a transition at $g_c=1$: for $g<g_c$ the network converges to a stable fixed point, while for $g>g_c$ the dynamics is chaotic.

The structure of the connectivity matrix determines whether the dynamics is relaxational. For generic (asymmetric) $J$, ~\eqref{eq:SCS_dynamics_methods} does not derive from an energy function, and the dynamics may exhibit chaotic behavior. However, when the couplings are symmetric, $J_{ij}=J_{ji}$, the dynamics becomes a gradient descent in an energy landscape,
\begin{equation}
\frac{\mathrm{d} h_i(t)}{\mathrm{d}t} = - \frac{\partial \mathcal{E}}{\partial h_i}\Big|_{h(t)},
\end{equation}
with energy
\begin{equation}
\begin{aligned}
\mathcal{E}(h_1,\dots,h_N) 
&= \sum_{i=1}^N \int_0^{h_i} \mathrm{d}h \, \phi(g h) \\
&\quad - \frac{1}{2}\sum_{i,j=1}^N J_{ij}\, \phi(g h_i)\,\phi(g h_j).
\end{aligned}
\label{eq:symmetrized_energy}
\end{equation}
The first term encodes the single-unit activation nonlinearity, while the second term represents pairwise synaptic interactions.

\subsection*{Binary excitatory--inhibitory model with symmetric interactions}

We also consider a binary-state network model in which each node is either active ($s_i(t)=1$) or inactive ($s_i(t)=0$) at discrete time $t$. Each node interacts with its $k$ neighbors through weighted inputs. The probability of activation is given by
\[
\mathcal{P}\big(s_i(t{+}1)=1\big) = f\!\left(\Lambda_i(t)\right),
\qquad
\Lambda_i(t)=\frac{\gamma}{k}\sum_{j}\omega_{ij}s_j(t),
\]
where $\gamma$ controls the global coupling strength and $\omega_{ij}\in\{\pm1\}$ encodes excitatory ($+1$) or inhibitory ($-1$) interactions. The activation function is taken to be piecewise linear:
\[
\begin{cases}
0, & \Lambda < 0,\\
\Lambda, & 0 \leq \Lambda \leq 1,\\[3pt]
1, & \Lambda > 1.
\end{cases}
\]

To connect directly with the symmetric interaction framework used in the present work, we consider a link-symmetrized variant in which inhibition and excitation are assigned to \emph{links} rather than nodes. The network is taken to be random-regular and undirected, so that each node has exactly $k$ neighbors and $\omega_{ij} = \omega_{ji} \in \{\pm 1\}$. A fraction $p$ of links are inhibitory and $(1{-}p)$ are excitatory, ensuring that the coupling matrix $(\omega_{ij})$ is symmetric by construction. The update rule remains unchanged, but inhibition is now encoded at the level of pairwise interactions rather than neuronal types, providing a minimal symmetric framework for a contact--process--like antagonistic coupling.}


%


\clearpage


\section*{Supplementary material}

# Supplementary Information: Topological Symmetry Breaking in Antagonistic Dynamics

Giulio Iannelli[1,2], Pablo Villegas[1,3],[*] Tommaso Gili[4], and Andrea Gabrielli[1,5]

[1]*Enrico Fermi Research Center (CREF), Via Panisperna 89A, 00184, Rome, Italy*
[2]*Dipartimento di Fisica, Università di Roma "Tor Vergata", 00133 Rome, Italy*
[3]*Instituto Carlos I de Física Teórica y Computacional, Universidad de Granada, Granada, Spain*
[4]*Networks Unit, IMT Scuola Alti Studi Lucca, Piazza San Francesco 15, 55100- Lucca, Italy. and*
[5]*Dipartimento di Ingegneria, Università degli Studi "Roma Tre", 00146, Rome, Italy*

**CONTENTS**

---

[*] pablo.villegas@cref.it

# I. RELATION BETWEEN THE GROUND STATE OF THE RBIM AND THE EIGENVECTORS OF THE SIGNED LAPLACIAN OPERATOR

Let the Random Bond Ising Model (RBIM) be defined by the Ising Hamiltonian

$$\mathcal{H}[\boldsymbol{\sigma}] = -\sum_{(i,j)} J_{ij}\sigma_i\sigma_i, \tag{1}$$

where $\sigma_i = \pm 1$ with $i = 1, 2, ..., N$ are the usual binary spin variables, $J_{ij}$ are independent random variables taking value $-J$ (with $J > 0$) with probability $0 < p < 1$ and value $+J$ with the complementary probability $1 - p$, and the sum $\sum_{(i,j)}$ extends to all pairs of first nearest neighboring spins. At $T = 0$ the system will assume the spin configuration, when it exists and is unique, that minimizes $\mathcal{H}[\boldsymbol{\sigma}]$. By noticing that

$$\sum_{i=1}^{N} k_i \sigma_i^2 = 2M,$$

where $M$ is the total number of interacting pairs ($M = dN$ in $d$-dimensional square lattices), we can rewrite Eq. (1) as

$$\mathcal{H}[\boldsymbol{\sigma}] = J \sum_{(i,j)} \bar{L}_{ij}\sigma_i\sigma_i - 2M,$$

Where $\bar{L}$ is the Signed Laplacian (SL) operator of the undirected graph of signed interactions. Thus minimizing $\mathcal{H}[\boldsymbol{\sigma}]$ is the same of minimizing

$$\mathcal{H}_{\bar{L}}[\boldsymbol{\sigma}] = \sum_{(i,j)} \bar{L}_{ij}\sigma_i\sigma_i$$

Note that when this configurations exists, its energy is lower bounded by the energy of the continuous field $\boldsymbol{\psi} = \{\psi_i\}$ minimizing the same quadratic Hamiltonian

$$\mathcal{H}_{\bar{L}}[\boldsymbol{\psi}] = \sum_{(i,j)} \bar{L}_{ij}\psi_i\psi_j$$

under the normalization condition $\sum_i \psi_i^2 = N$. In order to solve this problem we can use the Lagrange multipliers approach, i.e. minimizing

$$\mathcal{H}'_{\bar{L}}[\boldsymbol{\psi}] = \sum_{(i,j)} \bar{L}_{ij}\psi_i\psi_j - \lambda \sum_i \psi_i^2$$

and then finding the suitable values of the multiplier $\lambda$. By differentiating for $\psi_i$ the above expression we get the eigenvalue equation:

$$\sum_{i=1}^{N} \bar{L}_{ij}\psi_j = \lambda\psi_i,$$

i.e. a necessary condition for the minimization is that $\boldsymbol{\psi}$ is a normalizable eigenvector of the operator $\bar{L}$. Consequently, a necessary and sufficient condition for the minimization of $\mathcal{H}_{\bar{L}}[\boldsymbol{\psi}]$, under the condition of normalizability of the vector $\boldsymbol{\psi}$, is that it is the eigenvector of $\bar{L}$ corresponding to its minimal eigenvalue $\lambda_0$. Note that $\lambda_0 = 0$ in case of balanced signed networks and $\lambda_0 > 0$ if it is unbalanced.

A problem arises when, due to the spatial distribution of the negative interactions/defects, i.e. to the structure of the operator $\bar{L}$, such a minimal eigenvector is no more unique, but a set of degenerate eigenvectors appears sharing the same minimal eigenvalue of $\bar{L}$. The onset of this situation clearly depends on the value of the fraction $p$ of the randomly located $S$−defects and topologically corresponds to the percolation of diffusion defects/barriers that prevents connectivity between different region of the network giving rise to a topological segmentation of the system and therefore to the breaking of the ergodicity of the diffusion dynamics on the lattice/network. This critical value of $p$ obviously depends on the kind of considered lattice/network.

Finally, note that the same argument extends its validity to any random interaction matrix $J_{ij}$ including the Edwards–Anderson (EA) model which links the ground state of a spin-glass to the fundamental eigenvector of the SL operator.

## II. ENTROPY REGULARITY FOR THE SIGNED LAPLACIAN

We show that the Laplacian entropy, as defined in the main text, converges to zero as the timescale $t$ goes to infinity by using that the SL is strictly positive definite. As explicitly stated in Methods, the entropy can be expressed as the sum of the internal energy contribution and the free energy of the ensemble of states. It is easy to see that in the case where the (signed) Laplacian matrix is positive semidefinite, that is, the smallest eigenvalue is greater than or equal to zero, $\lambda_0 \geq 0$, it holds $S(t) \to 0$ for $t \to \infty$.

$$\lim_{t \to \infty} S(t) = \lim_{t \to \infty} \left[ \frac{\sum_i \lambda_i t e^{-\lambda_i t}}{Z(t)} + \ln Z(t) \right] =$$
$$= \lim_{t \to \infty} \left[ \frac{\sum_i \lambda_i t e^{-\lambda_i t}}{\sum_i \lambda_i e^{-\lambda_i t}} + \ln \left( \sum_i e^{-\lambda_i t} \right) \right] = 0$$

Hence, all $\lambda_i t e^{-\lambda_i t} \to 0$, since $\lambda_i > 0$ for $i > 1$, and $\lambda_0 = 0$. Instead, in the second term the only remaining component in the logarithm is $e^{-\lambda_0 t} = 1$. When $\lambda_0 > 0$ this feature is preserved, as can be seen if we gather a $e^{-\lambda_0 t}$ factor,

$$\lim_{t \to \infty} S(t) = \lim_{t \to \infty} \left[ \frac{\sum_i \lambda_i t e^{-(\lambda_i - \lambda_0)t}}{\sum_i e^{-(\lambda_i - \lambda_0)t}} - \lambda_0 t + \ln \left( \sum_i e^{-(\lambda_i - \lambda_0)t} \right) \right] =$$
$$= \lim_{t \to \infty} \left[ \frac{\sum_i (\lambda_i - \lambda_0) t e^{-(\lambda_i - \lambda_0)t}}{\sum_i e^{-(\lambda_i - \lambda_0)t}} + \ln \left( \sum_i e^{-(\lambda_i - \lambda_0)t} \right) \right] =$$
$$= \lim_{t \to \infty} \frac{\mathrm{d}}{\mathrm{d}t} \left[ t \ln \left( \sum_i e^{-(\lambda_i - \lambda_0)t} \right) \right] = \frac{\mathrm{d}}{\mathrm{d}t} \lim_{t \to \infty} \left[ t \cdot \frac{\ln \left( 1 + O(e^{-(\lambda_1 - \lambda_0)t}) \right)}{O(e^{-(\lambda_1 - \lambda_0)t})} \cdot O(e^{-(\lambda_1 - \lambda_0)t}) \right] = 0,$$

demonstrating that the entropy converges to zero as $t \to \infty$.

## III. PERCOLATION OF DEFECTS IN REGULAR LATTICES

To rigorously investigate the physical nature of the transition observed in the entropy and specific heat of the density matrices, we employed a systematic approach by progressively introducing negative links into the lattice topology. This allowed us to analyze the resulting dynamical effects on the field $\psi$, governed by the heat–like equation:

$$\partial_t \psi = -\bar{L}\psi. \tag{2}$$

Given that the long–term behavior of the system is dominated by the lowest eigenstate $|\lambda_0\rangle$ of the SL, where $\psi(t) = e^{-t\bar{L}}\psi(0)$ leads to $\psi(t \to \infty) = |\lambda_0\rangle$, our analysis focused on the properties of $|\lambda_0\rangle$ as the number of negative links increased. Initially, we observed a phenomenon closely resembling *Anderson localization* (see Fig. 1) of $|\lambda_0\rangle$, where most components are significantly smaller compared to a concentrated region that contains the bulk of the measure. This condition arises as the severe interference among multiple scattering paths effectively halting the diffusion process, trapping the wave components within the disordered medium.

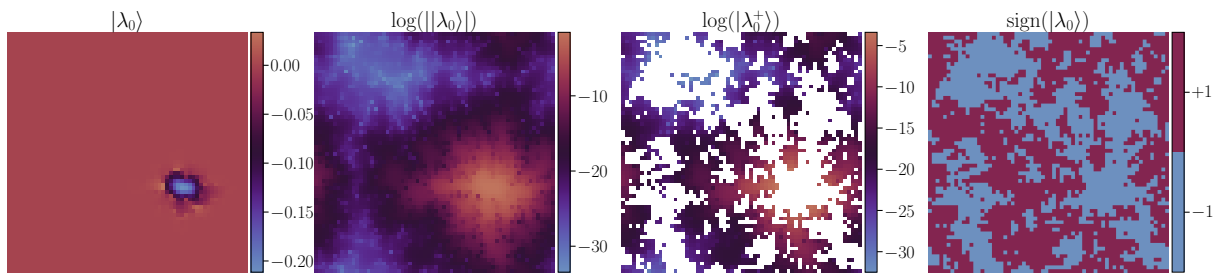

FIG. 1. Localization effects and signed structure of the lowest eigenvalue $|\lambda_0\rangle$ on a signed 2D square lattice when $p \approx p_\mathrm{c}$.

As the negative links increased, this localized region evolved, leading to a percolation–like phenomenon in the sign structure of $|\lambda_0\rangle$. At the critical threshold $p_\mathrm{c}$, binarizing the eigenvector by taking $\mathrm{sign}(|\lambda_0\rangle)$ permits to reveal the emergence of a giant component (rightmost plot of Fig. 1) when projected onto the lattice topology. This behavior suggests a percolative transition in the system.

## A. The analogy of topological defects as continuum percolation of disks

To further investigate the topological implications, we focused on analyzing the field $\psi$ in Eq. (2) when small perturbations, in the form of localized negative links, were introduced into the system. We refer to these basic "excitations" of the lattice topology as *topological defects*. As shown in Fig. 2, we classify the elementary structures as single (S), Z, and X defects. The Z defect represents the plaquette flip, where all links in the elementary geometric cell are negatively signed, and the X defect corresponds to antagonistic nodes, where all links connected to a single node are negatively signed. On the left side of the figure, these defects are depicted on a square lattice, while the right side shows analogous results for a triangular one. The first row illustrates the effect of a single S defect, the second row represents a single Z defect, and the third row a single X one, with system sizes ranging from $2^6$ to $2^{14}$ nodes. Each plot includes perpendicular and parallel cuts of the bidimensional field $\psi$, reflecting its behavior on both lattice types.

From this, we can discuss a preliminary picture: the X defect effectively isolates the antagonistic node from its neighbors, as evidenced by the negative component of the projected eigenstate onto the topology, which only corresponds to the "isolated node". Note that, as long as two X defects are not adjacent to each other, the graph remains balanced, and the lowest eigenstate can distinguish between nodes in the positive and negative partitions exactly. In contrast, Z defects induce local unsatisfied conflicts that produce a continuum of influence, decaying logarithmically towards the center of the plaquette. A single defect behaves similarly to a Z defect, depending on the lattice type.

The Laplacian operator model spatial diffusion processes, with its eigenfunctions representing plane waves in homogeneous systems. However, in the case of the *signed Laplacian*, additional complexity arises due to the inclusion of *negative links*, which we will show acting as localized *sinks*. These negative links deplete the field $\psi$ locally, modifying the diffusion process. Notably, when a graph is no longer balanced –meaning it contains negative links where local

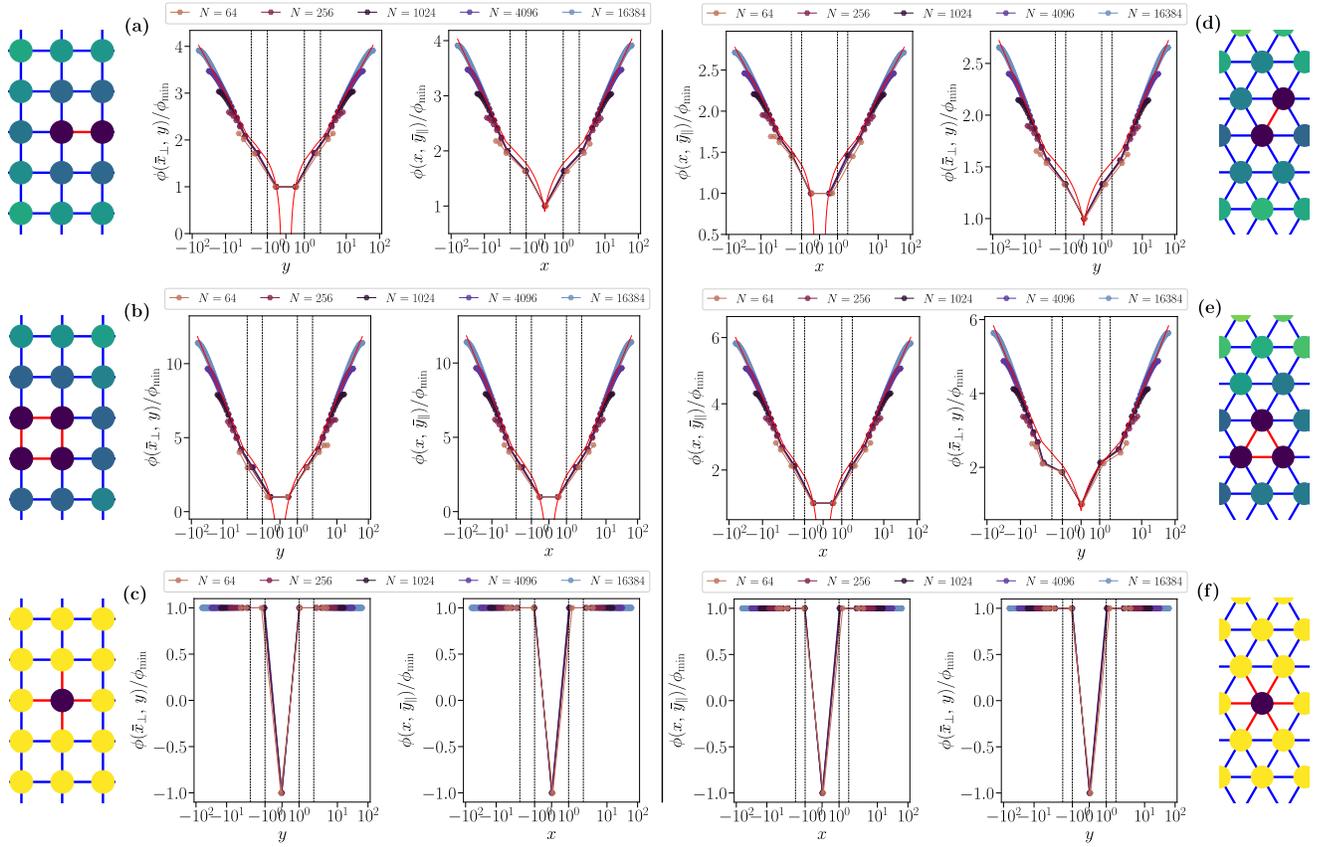

FIG. 2. Single, Z, and X defects on squared and triangular lattice. Fig. 2(a)–Fig. 2(c) refer to the three basic defects on the 2D square lattice, while Fig. 2(d)–Fig. 2(f) represent the same quantities but in for a triangular 2D lattice. For each one of the three defects (single, Z and X) the plot of the field in the tangent $\psi_\parallel$ and transverse $\psi \perp$ direction is shown. The distance on the spatial axis is in *symmetrical log scale* to properly show the logarithmic behavior of the density field at long distance, indicating the absence of a proper action such as

conflicts always remain unresolved (as in the case of unbalanced triangles)– the SL becomes *positive definite*, altering the behavior of the system. The interpretation of negative links as sinks arises from this positive definiteness: the SL introduces a depletion of the field, mimicking the behavior of sinks in a diffusion equation. These defects can also be understood as scatterers in two senses: first, as obstacles to the propagation of plane waves, which are the homogeneous solutions to the diffusion equation governed by the Laplacian, and second, as regions that distort the diffusion process in the presence of a field. The interplay of diffusion, plane waves, and the Laplacian is well–established, and in this context, embedded in the SL. Notice that for X defects the effect on diffusion is more similar to a *reflective* boundary, where no information is exchanged between its interior and its exterior.

We adopt a similar formalism to that presented in the analysis of diffusion with sinks [1], dealing with the SL process as analogous to diffusion, where negative links (excluding isolating defects like X defects) generate an "effective" influence sphere of a certain radius, affecting the surrounding region. This approach is explained in detail in the following considerations.

The SL on a discrete graph, as demonstrated in the main text, can be written as $\bar{L} = -L + S$ where $S$ is a correction term encoding the presence of negative links.

**a. Passage to Continuum and Helmholtz–Like Equation.** Considering the lattice spacing $a \to 0$ in the *continuum limit*, the discrete Laplacian converges to the continuous operator $\nabla^2$. This transition allows us to model the behavior of the SL as a diffusion process with a localized sink in the continuum case. For a single negative link located at the origin, $S$ can be modeled by a Heaviside function, $S = \kappa\Theta(r - R_{\text{snk}})$, where $R_{\text{snk}}$ defines the radius of the sink and $\kappa$ represents the strength of the negative link. The SL in this limit takes the form:

$$\bar{L} = -D\nabla^2 + \kappa\Theta(r - R_{\text{snk}}), \tag{3}$$

where $D$ is the diffusion constant. The Heaviside function $\Theta(r - R_{\text{snk}})$ activates the sink within a finite region, turning off its influence beyond $r = R_{\text{snk}}$. This describes a system where the negative links act like circular (in 2D) or spherical (in 3D) sinks, modifying the diffusion operator.

Now, we consider the heat–equation from Eq. (2), that can be used to find the steady state of the scalar field, $\psi$, by substituting the continuum SL and imposing stationarity,

$$\nabla^2 \psi = \frac{\kappa}{D}\Theta(r - R_{\text{snk}})\psi. \tag{4}$$

This equation can also be understood as analogous to a Helmholtz–like equation in the presence of scatterers, namely,

$$-D\nabla^2\psi + S\psi = \lambda\psi.$$

The presence of negative links modifies the eigenvalue structure of the SL, leading to the eigenfunctions as reported in Fig. 3(b). The passage to the continuum highlights how the SL acts in the long–wavelength limit, where the discrete lattice becomes dense, and the negative links correspond to regions that absorb the field rather than supporting its propagation. The solution to this equation in different dimensions captures the spatial behavior of the field in the presence of such sinks, corroborating the idea that the defects can be really modelled as local sinks. Then in order not to obtain null solutions we include an isotropic source proportional to the amount of density that the defect drain.

**b. Scalar field behavior in 2D.** Given the sink ansatz in the SL of Eq. (3) it is easy to solve now Eq. (4). For $r > R_{\text{snk}}$, where $\Theta(r - R_{\text{snk}}) = 0$, the last equation reduces to the purely radial Laplace equation, as the sink term $S$ vanishes. It follows

$$\psi(r) = C_1 \log(r) + C_2. \tag{5}$$

Thus, outside of the sink, the scalar field exhibits logarithmic decay as confirmed by numerical observations in Fig. 3(a). $R_{\text{snk}}$ will be thus the distance from the origin of the defect where the logarithmic scaling of the field breaks.

**c. Scalar field behavior in 3D.** As in the 2D case, the *angular components* of the Laplacian disappear due to the radial symmetry of the problem. For $r > R_{\text{snk}}$, the influence of the sink vanishes ($\Theta = 0$), and the solution is now

$$\psi(r) = \frac{C_1}{r} + C_2. \tag{6}$$

Again, plotting the effect of Z defect on the lattice on the density field $\psi$ yields the correct scaling of the field at $r > R_{\text{snk}}$.

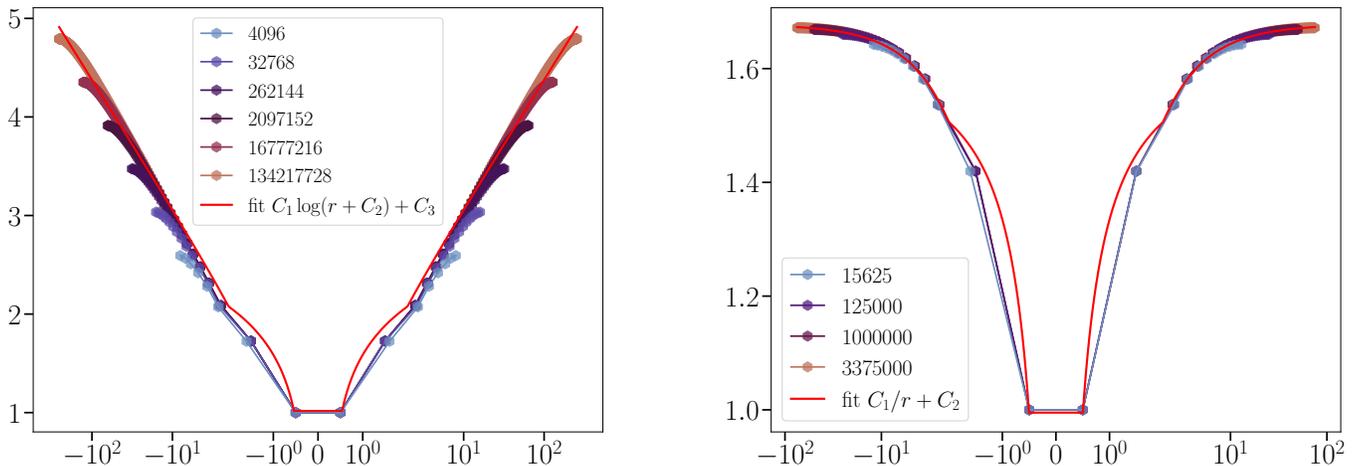

FIG. 3. As in Fig. 2: on the left the parallel cut w.r.t. to the signell defect on the 2D lattice, showing the logarithmic asymptotic behavior of the defect outside the region of influence. On the right the parallel cut of the field behavior in the case of a 3D simple cubic lattice,

**d. Considering multiple defects.** The above picture leads us to interpret the transition from a homogeneous to a disordered medium as one where these spherical defects percolate, forming a giant cluster in a continuum percolation process. In the case of Z defects, the system behaves in a way that is directly analogous to the classical continuum percolation problem. The emergence of a giant cluster of disks with radius $r$ in real space mirrors the critical behavior of continuum percolation, where the total area of the disks at the critical threshold is given by $\eta_c = \pi r^2 N / L^2$. Here, $r$ is the radius of the influence sphere, $N$ is the number of defects, and $L$ the system size. For single negative links, however, the situation slightly differs. In this case, the number of disks corresponds directly to the number of negative links, rather than to plaquettes with all edges flipped. This leads to the observation that the critical number of negative links scales as $N_{n,c}/L^2 \approx 2p_c$ for the square lattice, becoming exact in the limit $N \gg 1$. As a result, we introduce an effective critical fraction, $p_{c,\text{eff}}$, which accounts for this difference. The critical radius can then be computed as:

$$r_c = \sqrt{\frac{\eta_c}{\pi p_{c,\text{eff}}}}.\tag{7}$$

This framework explains how both Z defects and single negative links contribute to the percolation transition, with $p_{c,\text{eff}}$ representing an *effective fraction of negative units* relative to the system size.

The aforementioned analogy between continuum percolation on the lattice and the percolation in the reciprocal space has been smoothly drawn in the case of the square lattice due to its symmetry properties, but the argument can be thought similar on other kinds of topologies. The analogy between continuum percolation on the lattice and percolation in reciprocal space is readily apparent in the case of the square lattice due to its symmetry properties, and a similar argument may apply to other topologies. However, exploring the definition of an "effective radius" of a percolating disk in exotic topologies is beyond the scope of this work and serves us to give a physical interpretation that is in full agreement with the order parameters in k-space proposed in the main manuscript. Furthermore, such investigations are of limited relevance as the primary objective of determining this radius is to deepen our understanding of the physical nature of defects.

### B. Percolation of S defects

To better elucidate and strengthen the analogy between the onset of disorder in the system and the corresponding percolation of topological defects, we have employed the critical radius Eq. (7) to demonstrate that the mapping between these two phenomena holds.

In Fig. 4, we show three configurations of defects on a lattice of size $96 \times 96$ for progressively increasing values of $p$, specifically $p \in \{0.05, 0.1, 0.25\}$. These values are selected to represent the three relevant cases $p \ll p_c$, $p \approx p_c$, and $p \gg p_c$. Each subfigure depicts, on the left, the binarized fundamental eigenstate $|\lambda_0\rangle$, projected onto the lattice, while on the right, the configuration of negative links is illustrated: each negative link is represented by a (black) disk of


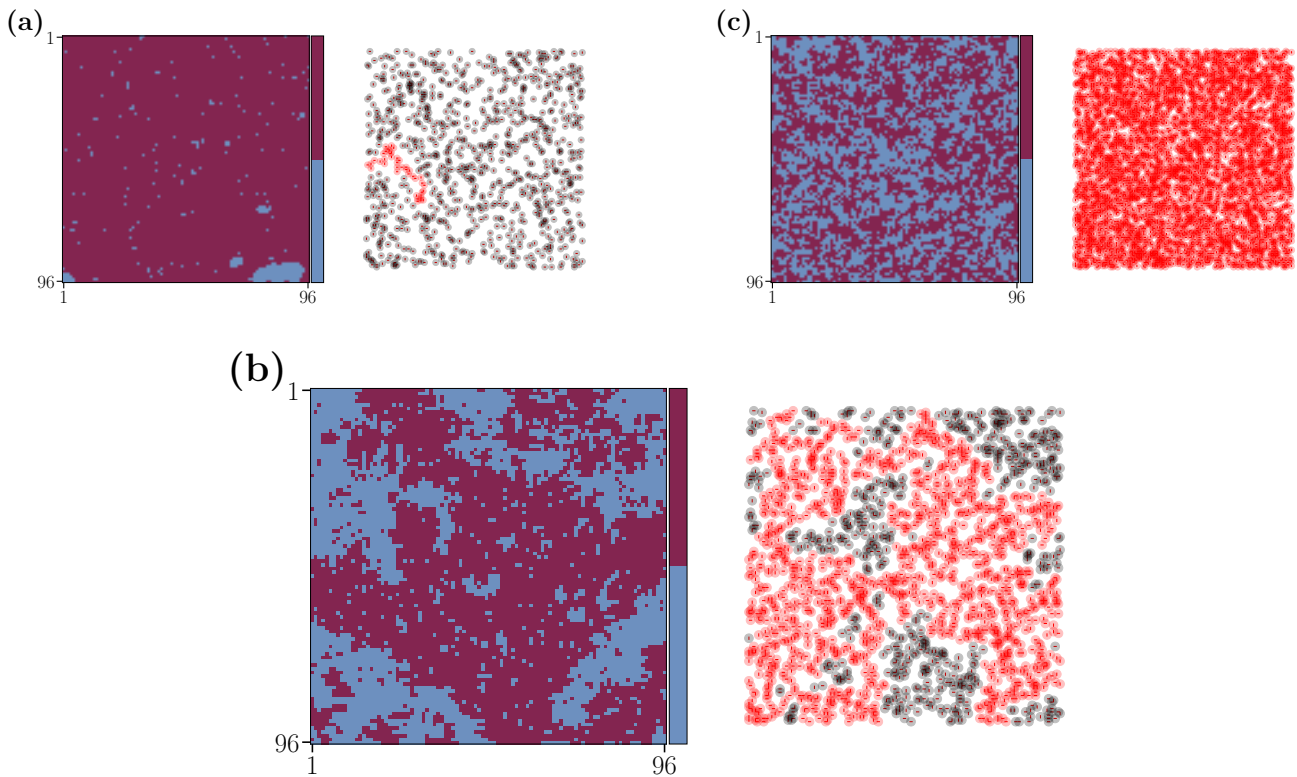

FIG. 4. **Percolation of S defects in 2D square lattices.** (Left) Snapshots of the smallest SL eigenvector $|\lambda_0\rangle$ projection for different defect fractions: **(a)** $p = 0.05$, **(b)** $p = 0.1$, and **(c)** $p = 0.25$. (Right) Sketch of the corresponding defects. Circles represent the critical percolating radius derived earlier. Note that a fractal percolating cluster emerges for $p = p_c$, both in reciprocal space, as seen in the components of the associated eigenvector, and in real space, as observed in the overlapping circles of radius $r_c$ (colored in red).

radius $r_c$. The largest percolating cluster of overlapping circles is highlighted in red. It is important to emphasize that there is no immediate connection between the spatial distribution of links and the spatial arrangement of the emerging giant cluster in the binarized lowest eigenstate. The physical analogy is established by the emergence, at the critical value $p_c$, of a giant cluster, which appears both among the elements of the lowest eigenstate with identical signs and in the disk percolation process characterized by radius $r_c$. This result reinforces the parallelism between topological phase transitions induced by the insertion of negative links and the percolation phenomenon. For completeness, and to provide a more comprehensive understanding of the universality of the observed phase transitions, we have extended the simulations discussed in the main text to additional lattice structures, including hexagonal (see Fig. 5), triangular (see Fig. 6), and 4D lattices (see Fig. 7). As discussed in the main text, the case of S defects can be exactly mapped onto a scenario that has been extensively studied in the literature in the context of the $\pm J$ Ising spin glass model at zero temperature, also known as the RBIM. Specifically, this corresponds to an EA model with a bimodal distribution of couplings. We therefore investigated whether the critical behavior and associated markers, such as the giant component $P_\infty$ and its corresponding fluctuations $\chi$, remain consistent across varying topological configurations and align with what is known about phase transitions in dynamical statistical systems, using only a geometric explanation. As we will show in the following figures, these different lattice types exhibit similar critical behavior, with clear indicators of the phase transition occurring near corresponding $p_c$, in agreement with previous studies [2], which identified this as the point where the growth rate of *frustrated plaquettes* equaled the growth rate of antiferromagnetic bonds. A *frustrated plaquette* refers to a closed loop or face in a lattice model where it is impossible to satisfy all the interaction bonds simultaneously due to the configuration of the couplings, namely where

$$\varsigma = \prod_{\langle i,j \rangle \in \square} \text{sign}(J_{ij}) \tag{8}$$

is negative, where the product is taken over all bonds $\langle i, j \rangle$ around the plaquette (smallest cycle in lattice), and $\text{sign}(J_{ij})$ represents the sign of the coupling $J_{ij}$ between the spins $i$ and $j$.

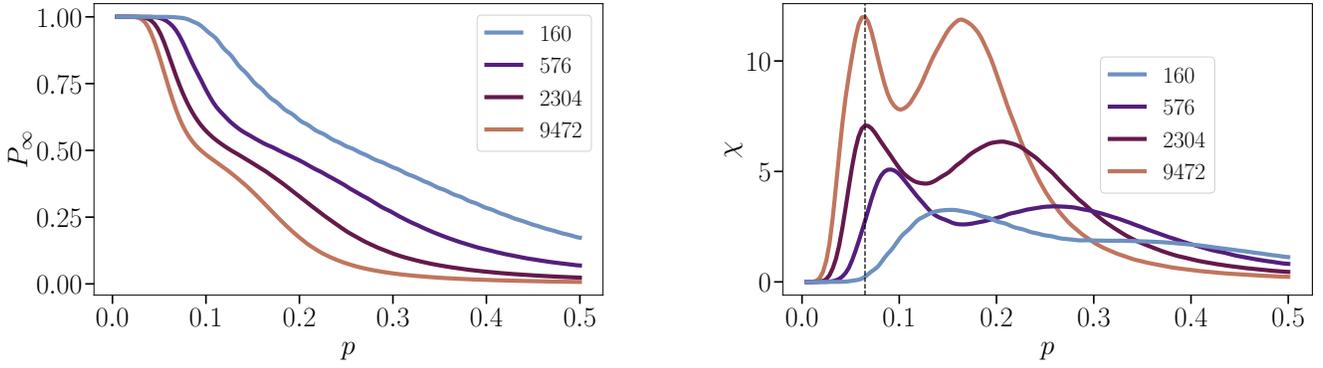

FIG. 5. **Hexagonal planar lattice order parameter and flutuations.** The order parameter $P_\infty$ is showed in **(a)** for an hexagonal lattice with sizes $(L_1, L_2) \in \{(16, 10), (32, 18), (64, 36), (128, 74)\}$. Subfigure **(b)** shows the corresponding susceptibilities. The double peak gap in the susceptibilities tends to close as $N \to \infty$ indicating a finite–size effect due to the dimensionality of the problem.

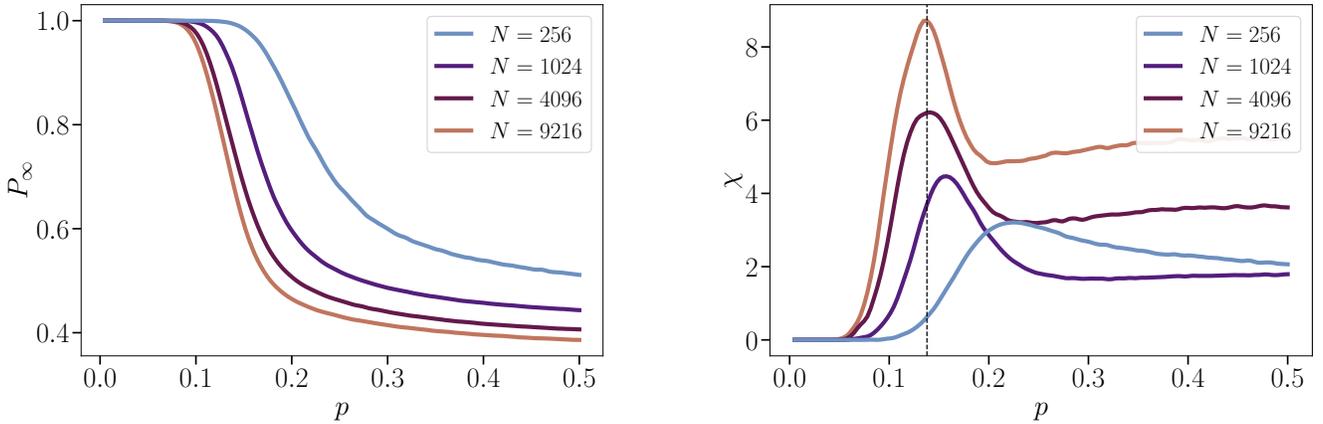

FIG. 6. **Triangular planar lattice order parameter and fluctuations.** The order parameter $P_\infty$ is shown in **(a)** for a triangular lattice with sizes $L \in \{16, 32, 64, 96\}$. The increased number of nearest neighbors (6, compared to 3 in hexagonal and 4 in square lattices) allows the triangular structure to avoid the finite-size effects seen in square and hexagonal structures, making its overall behavior less susceptible to the artifacts that arise from finite-size limitations. However, the fluctuations do not vanish after the transition; rather, the system remains in a geometrically frustrated state. This frustration is inherent to the triangular geometry, as the arrangement of interactions prevents all local interactions from being simultaneously minimized.

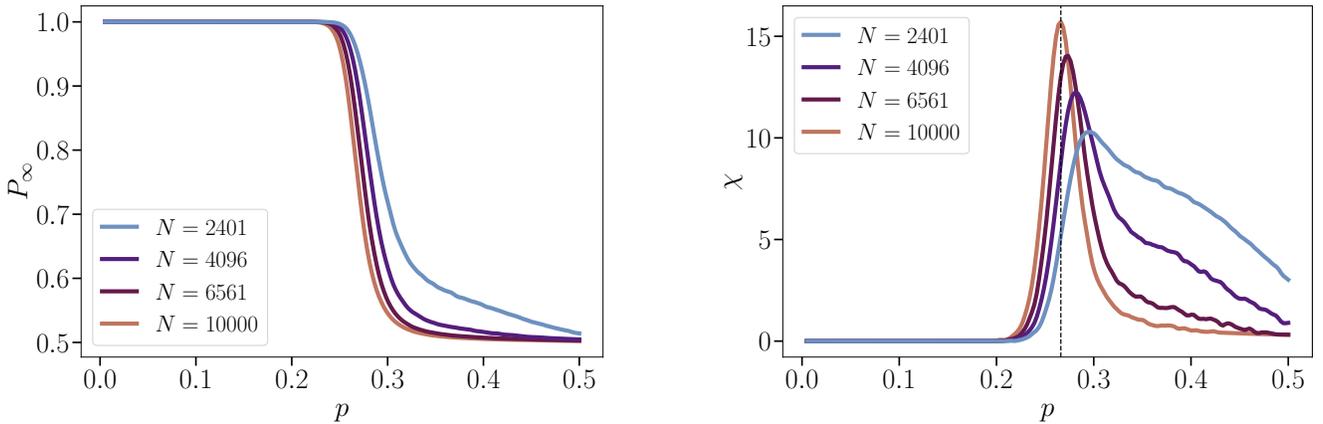

FIG. 7. **4D lattice order parameter and fluctuations.** The order parameter $P_\infty$ is shown in **(a)** for a 4D lattice with sizes $L \in \{7, 8, 9, 10\}$. Unlike the 2D and 3D cases, the 4D lattice exhibits a single, sharp peak in the order parameter, which indicates mean-field-like behavior typical of higher-dimensional systems (except for $L = 7$ where significant finite-size effect occur).

The values obtained for $p_c$ are comparable to the ones presented in geometrical studies as well as those with finite size scaling on the RBIM: $p_{c,\text{hex}} \sim 0.064(7)$, $p_{c,\text{tri}} \approx 0.1 - 0.15$, $p_{c,\text{4D}} \approx 0.28(1)$ [3, 4].

In 2D square and hexagonal lattices, susceptibility shows two distinct peaks. As the system size grows, these peaks come closer together. The first peak represents localized frustration in the structure, while the second marks the formation of a global percolating cluster. Eventually, these peaks merge, suggesting that the system reduces its local disorder as it grows. On the other hand, triangular planar lattices exhibit a single peak followed by a constant susceptibility. This likely results from the increased connectivity in triangular structures. At the same time the triangular cell prevents complete resolution of frustrations (the flipped version of the basic unit structure is frustrated, i.e. $\varsigma = -$ as opposed to square and hexagonal cells), leaving some inherent fluctuations in the system. In 3D cubic lattices, the double–peak persists even in large systems. This suggests a characteristic feature of percolation dynamics in three dimensions, possibly tied to the coexistence of multiple spanning clusters and consistent with the Alexander-Orbach conjecture (see Main). The double peak reflects this interplay between local frustration and large-scale connectivity. For 4D lattices, the susceptibility shows a single sharp peak, suggesting mean-field behavior. The increased dimensionality reduces the role of local fluctuations, leading to a more uniform percolation transition and aligning with expectations.

### C. Percolation of Z defects

In a very similar fashion to what was done previously for the single defect case, we here report the analogous figures for the Z defect and the corresponding phase transition. In this case, by $p$ we are counting the fraction of squared cells that are selected for flipping all the links associated with them. Thus, $p = 0.1$ in this context means that 10% of the plaquettes have been flipped.

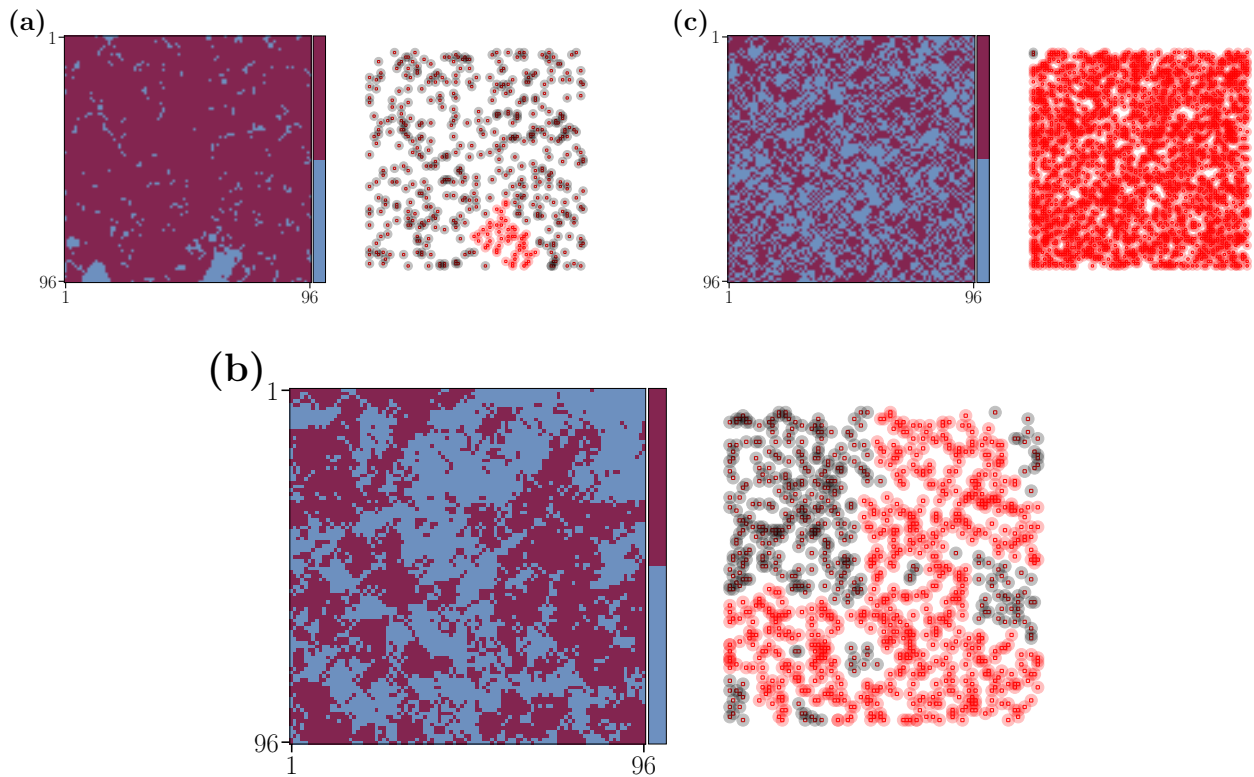

FIG. 8. **Percolation of Z defects in 2D square lattices.** Snapshots of the projected smallest eigenvector $|\lambda_0\rangle$ for different defect fractions: **(a)** $p = 0.05$, **(b)** $p = 0.1$, and **(c)** $p = 0.25$. On the left of each subfigure, we show the binarized fundamental eigenstate projected onto the lattice, while on the right, we depict the configuration of negative links. The circles represent the critical percolating radius derived earlier. The largest percolating cluster of overlapping circles is highlighted in red, indicating the emergence of a giant cluster for $p = p_{\text{crit}}$.

As can be observed from Fig. 8, the emergence of the giant cluster both in the spectrum and in real space occurs at approximately the same value of $p$ as the edge flipping in the S defect case. This demonstrates a similar percolation

threshold, reinforcing the notion that these Z defects are effectively the minimal geometrically frustratable units. Interestingly, the equivalence between flipping entire squares (plaquettes) and flipping individual links shows that the critical behavior is not dependent on the specific nature of the perturbation, but rather on the presence of disorder itself. This points towards a form of universality: the percolation transition appears to be driven by the overall density of frustrated bonds or defects, irrespective of whether the defects are introduced through flipping individual links or complete cells. Thus, the critical percolation properties observed in both cases suggest that the system's behavior is robust and governed by an underlying universal mechanism that transcends the specifics of individual versus collective perturbations.

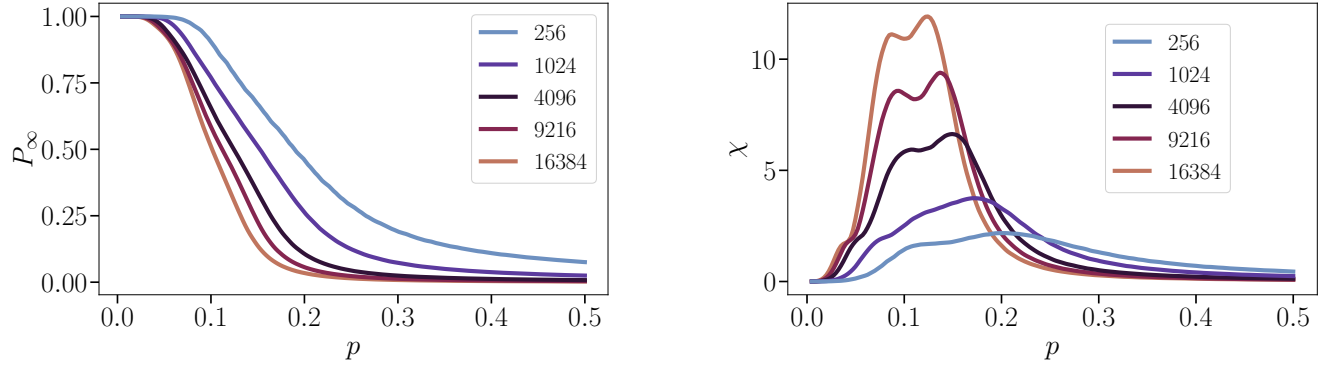

FIG. 9. **Percolation of Z defects in square lattices.** The order parameter $P_\infty$ is shown in **(a)**. Subfigure **(b)** depicts the corresponding susceptibility, showing the phase transition behavior as $p$ increases.

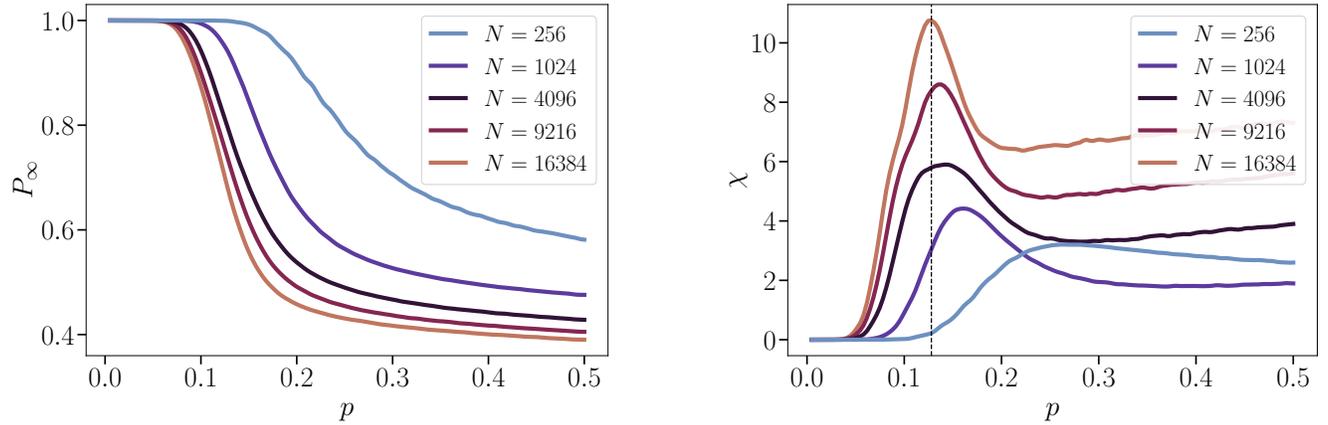

FIG. 10. **Percolation of Z defects in triangular lattices.** The order parameter $P_\infty$ is shown in **(a)** for a triangular lattice with different system sizes. Subfigure **(b)** illustrates the corresponding fluctuations, highlighting the differences in percolation behavior compared to square lattices due to the increased connectivity in triangular structures.

### D. Percolation of X defects

Analogously to the case of Z defects, we point out that its nodes "getting flipped" and not edges. This means that the fraction of flipped entities, $p$, represents the nodes that are only negatively connected to other nodes. Nonetheless, in this case, the scenario is significantly different compared to S and Z defects. An X defect involves flipping all the links connected to a particular node, effectively isolating it from the network. In our analogy, X defects behave like reflective barriers, in contrast to S and Z defects (see Fig. 2), which act like sinks where information/measure is drained. With X defect, no information can travel across the affected node, resulting in local isolation. Unless two of them touch. Initially, as X defects are introduced into the lattice, they do not significantly unbalance the network unless two X defects are sufficiently close to each other. When two X defects meet they create paths of frustration through the system. This interaction leads to two distinct transitions: Fig. 11 and Eq. (12) exhibit indeed

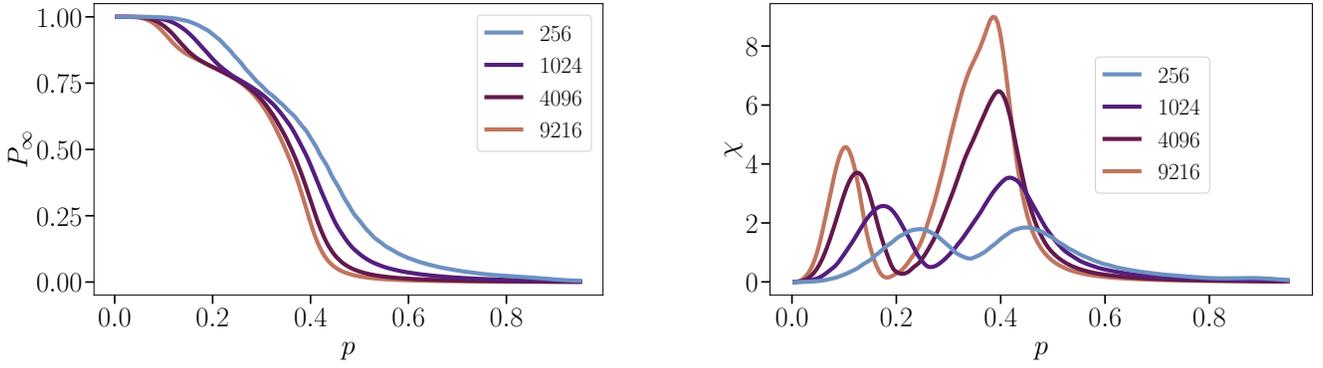

FIG. 11. **Percolation of X defects in squared lattices.** The order parameter $P_\infty$ is shown in **(a)** for a squared lattice with different system sizes and a fraction $p$ of X defects. Subfigure **(b)** illustrates the corresponding fluctuations. The double peak structure highlights the presence of the intermediate multistable phase.

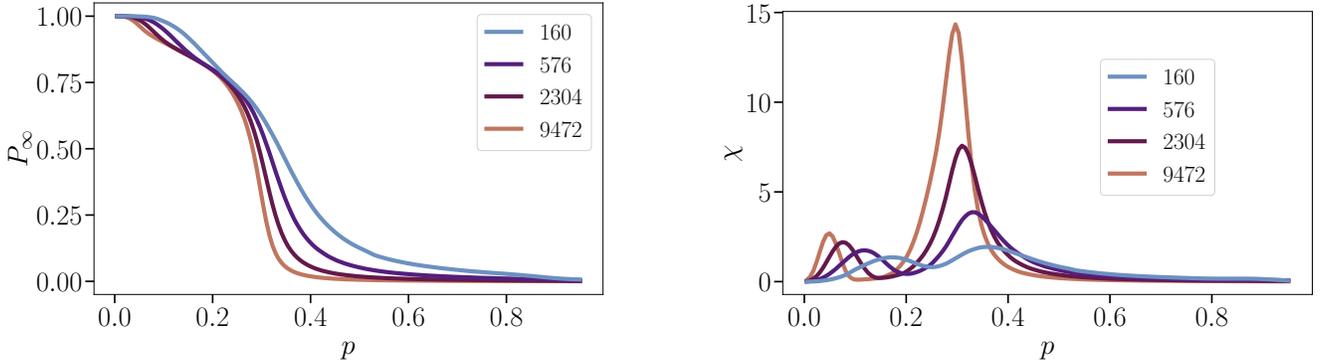

FIG. 12. **Percolation of X defects in hexagonal lattices.** The order parameter $P_\infty$ is shown in **(a)** for an hexagonal lattice with different system sizes and a fraction $p$ of X defects. Subfigure **(b)** illustrates the corresponding fluctuations. The double peak structure highlights the presence of the intermediate multistable phase.

the emergence of a double peak structure in the susceptibility $\chi$. The first peak is associated to the fragmentation of the macro scale order, while the second to the breaking of the microscale order, aka the Spin Glass (SG) transition.

The first transition can be interpreted as the beginning of the fragmentation of the lattice modes into isolated regions, where X defects serve as barriers preventing the free propagation of the field. The second transition occurs at a higher value of $p$, where the competition between the emergent antiferromagnetic and ferromagnetic behaviors becomes evident. This phase transition signifies the formation of antiferromagnetic order in the system. The critical points for the two transitions are distinct: the first marks the point at which local clusters of X defects begin to form and connect, while the second is where a giant cluster forms, encompassing a significant fraction of the system. The presence of these two critical behaviors highlights the complex interplay between local isolation and global connectivity, resulting in a competition between different types of order. Thus, X defects introduce a unique mechanism into the system, different from the single and Z defect cases. Instead of allowing information to flow or creating local sinks, X defects reflect and isolate, ultimately giving rise to a rich set of phase behaviors involving multiple competing orders.

In Fig. 13, we show the configurations of defects on a lattice of size $96 \times 96$ for progressively increasing values of $p$, specifically $p \in \{0.01, 0.25, 0.5\}$. These values are chosen to illustrate the two distinct transitions mentioned above. Each subfigure depicts, on the left, the binarized fundamental eigenstate $|\lambda_0\rangle$, projected onto the lattice, while on the right, the configuration of X defects is illustrated. The formation of isolated clusters, followed by the onset of a giant cluster, can be clearly observed.

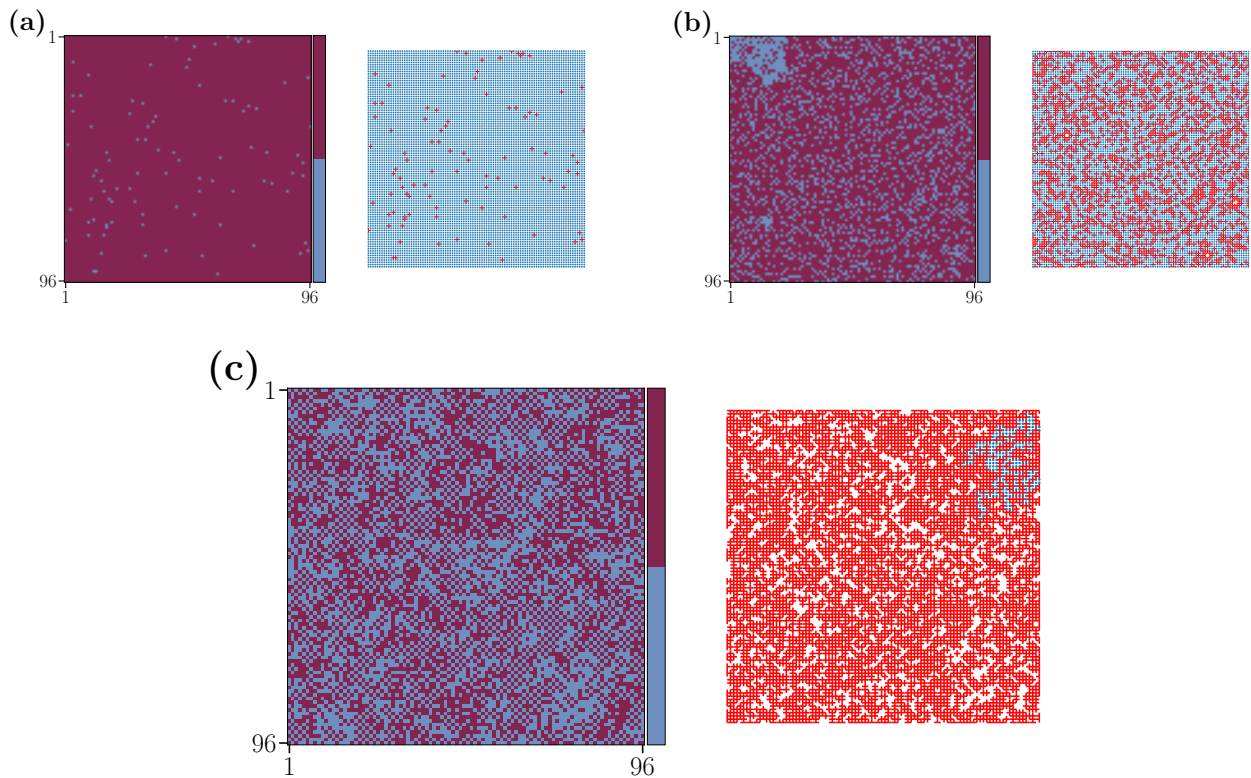

FIG. 13. **Percolation of X defects.** (Left) Snapshots of the projected smallest eigenvector $|\lambda_0\rangle$ on the lattice for different defects fraction: **(a)** $p = 0.1$, **(b)** $p = 0.25$, and **(c)** $p_c = 0.5$. (Right) Sketch of the corresponding defects. The X defects are represented as red crosses involving flipped nodes, while the blue dots are the nodes belonging to the GC of the positively connected ones. After the second peak transition this GC does not span the system anymore and ferromagnetic islands compete with antiferromagnetic ones contributing to the emergence of local to macroscopic disordered behaviors.

## IV. NUMERICAL ANALYSIS

### A. Resolution limits

The above analysis is subject to intrinsic numerical issues since it relies on computational tools for working out the SL spectrum. Particularly, the case of the lattices is illustrative of the actual numerical problems for large system sizes. As illustrated in Fig. 14, it can be noticed how the entries of the eigenstate(s) quickly become smaller than the smallest difference between two numbers that the machine can represent. A possible solution would be, therefore, to increase the floating point representation depth. Nevertheless, at double–precision floating–point format (i.e., the common 64–bit representation system available on usual machines), there is a significant fraction of *unreliable* values for square lattices at the order of $N \sim 10^4$ nodes.

The resolution limit is reached by both increasing the size of the system and increasing the number of defects in the topology. We have carefully ensured that this limit is not reached in all the results presented in this work.

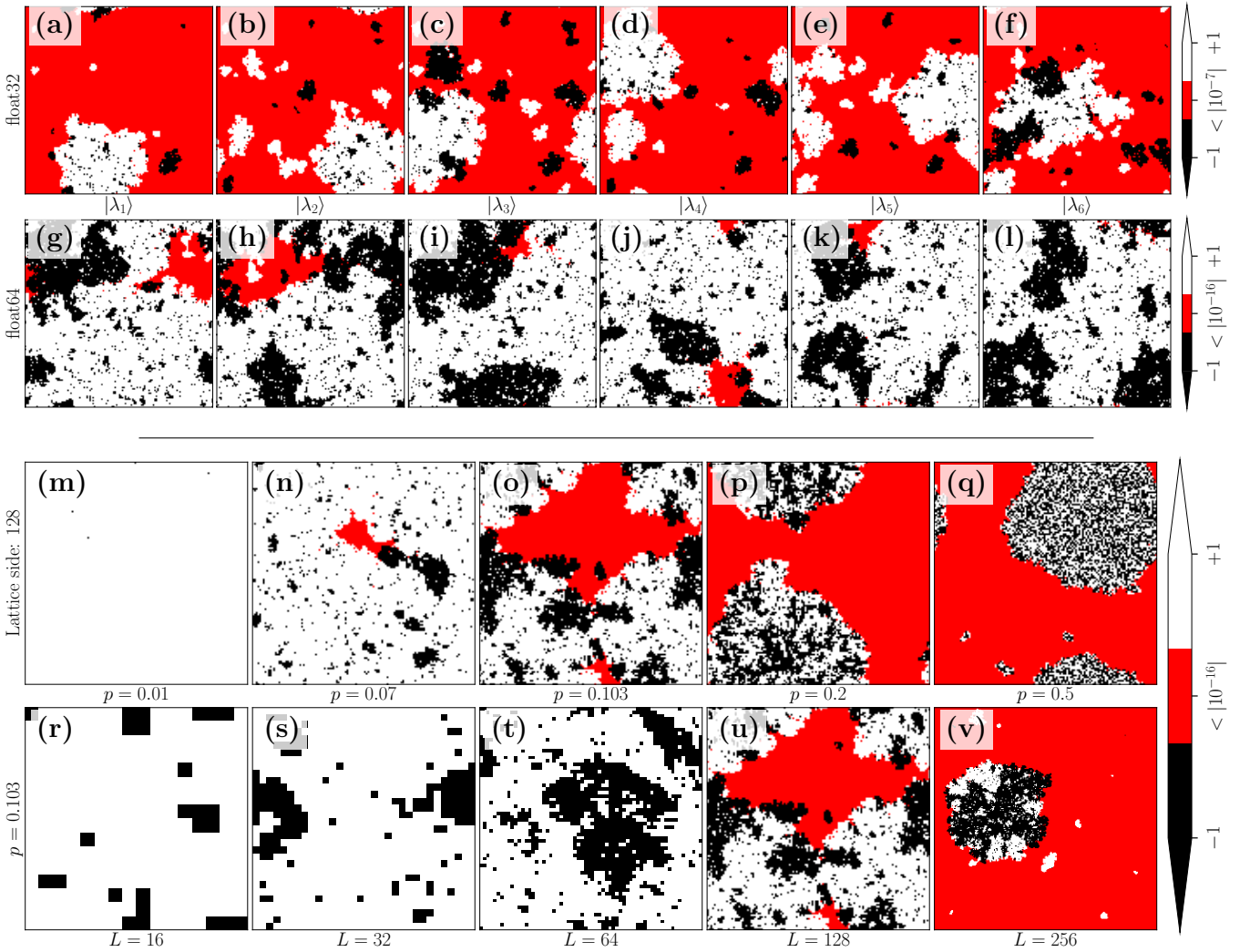

FIG. 14. Resolution limit constraints. **(a)**–**(f)** Binarized first six eigenstates of the SL computed with 32–bit (single precision) floating point arithmetic at $p = 0.103$ on a $128 \times 128$ square grid. Most of the values in this case already fall out of the resolution limit ($10^{-7}$). Doubling the precision, plots **(g)**-**(l)**, is a partial fix of this issue, but yet it does not allow for reliable clustering analyses. Plots **(m)**–**(q)** show how the number of unrealiable values varies with growing $p$, and analogously plots **(r)**–**(v)** with increasing side $L$.

### B. Signed Laplacian spectral probability distributions

The distribution of the values of the lowest eigenstates are reported in Eq. (15) for a planar squared lattice for four different values of the disorder parameter $p$. First it is worth mentioning that the negative peaks (corresponding to the negative entries of $|\lambda_0\rangle$) become more and more relevant in the distribution with increasing $p$. This is the mechanism leading to the topological phase transition.

As it is possible to observe with growing lattice size the numerical algorithm used for the computation (i.e. the Lanczos algorithm) of the latter suffers of a resolution limit where more and more values fall in the range $> |10^{-16}|$ which is approximately the machine precision. For bigger sizes there is a peak in these values which cannot be considered reliable anymore, at least for computing the giant cluster. Yet these value do not affect the cumulative distribution where they count infinitesimally.

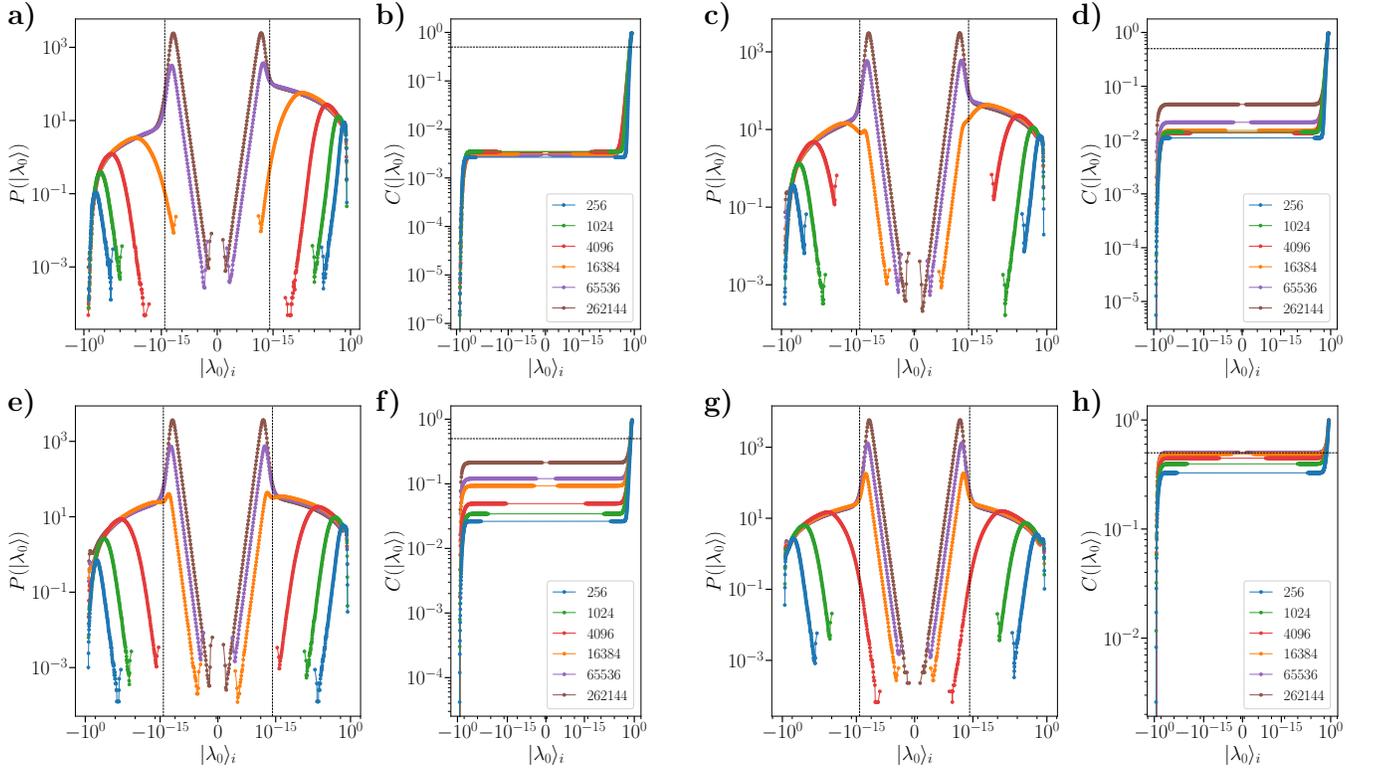

FIG. 15. **First eigenstate empirical distribution for the squared lattice.** Plots **(a)**, **(c)**, **(e)**, **(g)** report the distribution $P(|\lambda_0\rangle)$ for $p \in \{0.06, 0.844, 0.103, 0.225\}$, while **(b)**, **(d)**, **(f)**, **(h)** the corresponding cumulative $C(|\lambda_0\rangle)$.

## V. TOPOLOGICAL SYMMETRY BREAKING IN THE SPECTRUM OF THE SIGNED LATTICE

Another hallmark of the topological phase transition described in the main text is evident from the distribution of the eigenvalues of the SL operator in 2D squared lattices, as shown in Fig. 16. When $p = 0$, the spectrum exhibits a pronounced singularity at $\lambda = 4$ (the so-called *van Hove singularity*), which originates from the saddle point in the dispersion relation of the standard Laplacian on a square lattice. This singularity reflects the accumulation of modes associated with the four nearest–neighbor diffusion processes. As $p$ increases, the introduction of negative bonds

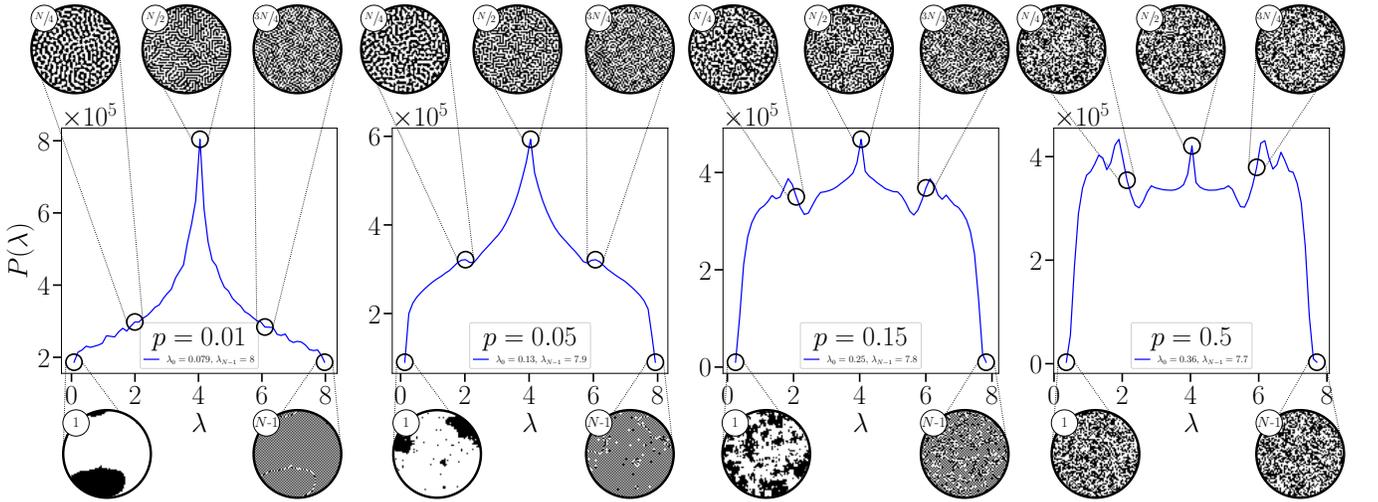

FIG. 16. Spectral distribution of the SL operator for bidimensional square lattices with increasing disorder $p$.

perturbs the regular structure of the lattice, and the spectral distribution begins to change. For $p < p_c$, the spectrum is still largely reminiscent of the $p = 0$ case; however, two lateral peaks start to emerge on either side of the main peak. These lateral features signal the onset of the topological symmetry–breaking process whereby the degeneracy among the diffusion modes is partially lifted.

For $p > p_c$, the lateral peaks become comparable to the fundamental one, a sign of the symmetry breaking of the system that is reflected in the spectral domain. At the same time, the correlation length of the spatial patterns directly depends on $p$: at low $p$, the lower system eigenstates correspond to smooth, large-scale diffusion modes, while at higher $p$ they become much more intricate and exhibit fractal features. Note that, near the critical regime, the lower eigenmodes also acquire a fractal structure, where the giant percolating cluster described in the main text emerges. Moreover, when approaching $p \to 0.5$, visual inspection of the spatial patterns clarifies how the characteristic correlation length of the patterns is comparable between them.

A particularly intriguing aspect of this evolution is that the emerging lateral van Hove–like singularities in $P(\lambda)$ resemble the overlap distribution function $P(q)$ encountered in Replica Symmetry Breaking (RSB) scenarios. In SG systems and related models, $P(q)$ develops a multi–peaked structure (or even a continuous distribution) as the system transitions into a state with many competing metastable configurations. Analogously, the appearance of lateral peaks in $P(\lambda)$ is directly related with the 'onset of degeneracy' presented in the main text.

## VI. DYNAMICAL EVOLUTION OF X DEFECTS

The phase diagram of the Ising dynamics on the square 2D lattice with X defects revealed an intermediate phase between the ferromagnetic regime and a glassy–like one. This regime is associated with a *multistable* phase where multiple temperature-resilient states emerge. Indeed, in this case, the X defects dynamically serve as topologically stable barriers, enhancing surface tension and allowing for domain walls to easily form and remain stable with respect to the inclusion of S defects. The metastable states can thus be interpreted as *memories*: in this analogy, if the system is initialized close in phase space to such states, they present a stable attraction basis preventing it from escaping, represented by the system eigenstate, $|\lambda_i\rangle$. This can be straightforwardly seen in Fig. 17 where a 2D square lattice is initialized with increasing SL eigenmodes (particularly $|\lambda_1\rangle, |\lambda_3\rangle, |\lambda_{10}\rangle, |\lambda_{50}\rangle$) and a random uniform configuration for comparison, performing a sudden quenching at $T = 0.5$. Aside from the lowest fully stable eigenmode, one observes over long Monte Carlo simulation times, that lower eigenmodes are indeed "close" to metastable configurations, which remain stable, while high-energy modes rapidly start to perform attractor surfing, evolving to other stable states that are expected to be linear combinations of the lower ones.

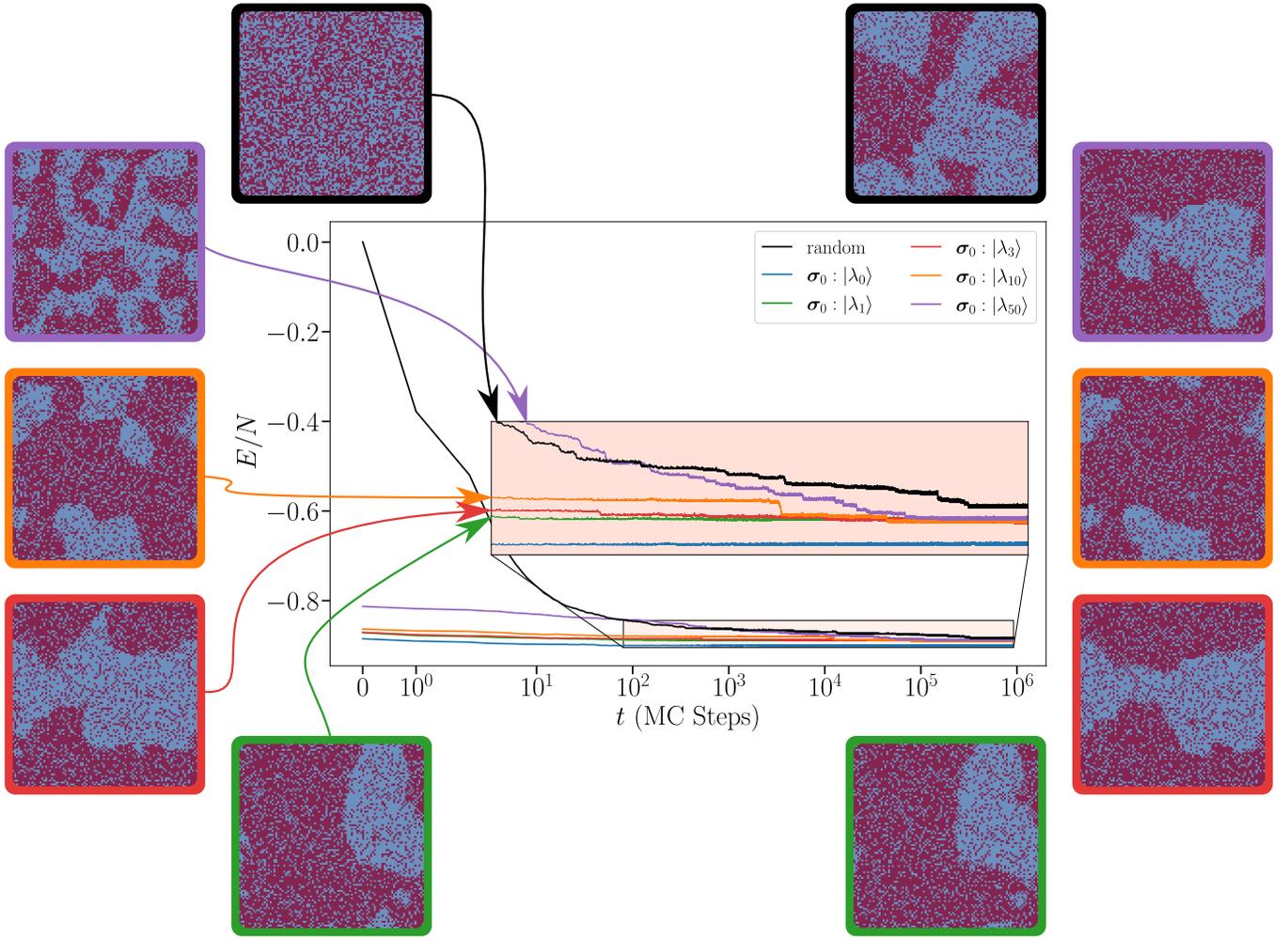

FIG. 17. Temporal evolution of different eigenstates of the system using X defects and considering a 2D squared lattice for Ising Dynamics in the metastable phase $p = 0.25$, $N = 96 \times 96 = 9126$, $T = 0.5$.



\end{document}